\documentclass[11pt]{article}

\usepackage{varioref}
\usepackage{amsmath,amsfonts,amsthm,mathrsfs}

\usepackage[normalem]{ulem}
\usepackage{multirow,color,graphics}
\usepackage{amsmath}
\usepackage{amsfonts}
\usepackage{amssymb}
\usepackage{jheppub}
\usepackage{enumerate}
\usepackage{fancyvrb}
\usepackage{verbatim}
\usepackage{wrapfig}
\usepackage{appendix}
\usepackage{amstext}
\usepackage{amssymb}
\usepackage{graphicx}
\usepackage{color}
\usepackage{varioref}
\usepackage{multirow,graphics}
\usepackage{epstopdf}
\newcommand{\nn}{\nonumber}

\numberwithin{equation}{section}

\def\[{\left[}
\def\]{\right]}
\def\({\left(}
\def\){\right)}
\def\d{\partial}

    \newcommand{\beq}{\begin{equation}}
    \newcommand{\eeq}{\end{equation}}
    \newcommand\beqa{\begin{eqnarray}}
    \newcommand\eeqa{\end{eqnarray}}
\newcommand\bea{\begin{array}}
\newcommand\eea{\end{array}}

\newcommand{\eq}[1]{(\ref{#1})}

\makeatletter
     \@ifundefined{usebibtex}{} {}
\makeatother

\newcommand{\bra}[1]{\langle #1 |}
\newcommand{\ket}[1]{| #1 \rangle}
\newcommand{\cN}{\mathcal{N}}
\newcommand{\mC}{\mathbb{C}}
\newcommand{\cH}{\mathcal{H}}
\newcommand{\rx}{\mathrm{x}}

\title{New Construction of Eigenstates and Separation of Variables for $SU(N)$ Quantum Spin Chains}

\author[a,b]{~~Nikolay Gromov,}
\author[c,1]{~~Fedor Levkovich-Maslyuk,\note{Alternative email: fedor.levkovich.maslyuk$\bullet$su.se}}
\author[d]{~~Grigory Sizov}

\affiliation[a]{Mathematics Department, King's College London,
The Strand, London WC2R 2LS, UK}
\affiliation[b]{St.Petersburg INP, Gatchina, 188 300, St.Petersburg,
  Russia}
\affiliation[c]{Nordita,
KTH Royal Institute of Technology and Stockholm University,\\
Roslagstullsbacken 23, SE-106 91 Stockholm, Sweden}
\affiliation[d]{LPT, Ecole  Normale  Superieure,  24  rue  Lhomond  75005  Paris,  France\\ }

\emailAdd{nikgromov$\bullet$gmail.com}
\emailAdd{fedor.levkovich$\bullet$gmail.com}
\emailAdd{grisha.sizov$\bullet$gmail.com}

\abstract{We conjecture a new way to construct eigenstates
 of integrable XXX quantum spin chains with $SU(N)$ symmetry.
The states are built by repeatedly acting on
the vacuum with a single operator $B^{\rm good}(u)$ evaluated at
the Bethe roots. Our proposal serves as a compact
alternative to the usual nested algebraic Bethe ansatz.
Furthermore, the roots of this operator
give the separated variables of the model, explicitly generalizing
Sklyanin's approach to the $SU(N)$ case.
We present many tests of the conjecture and prove it in several special cases. We focus on rational spin chains with fundamental representation at each site, but expect many of the results to be valid more generally.
}

\preprint{NORDITA-2016-111}

\begin{document}

\maketitle

\newpage

\section{Introduction}


Integrable quantum spin chains are among the most famous exactly solvable models in mathematical physics. In addition to exhibiting rich mathematical structures, they have found diverse applications ranging from condensed matter physics to supersymmetric gauge theories and string theory. The key feature of these spin chains is a powerful hidden symmetry which in particular allows one to obtain eigenvalues of the spin chain Hamiltonian in terms of the Bethe ansatz equations \cite{Bethe:1931hc} (see e.g \cite{Faddeev,Sutherland,Hubbook,Kundu:1996hb,Bombardelli:2016rwb} for pedagogical reviews). At the same time, the problem of efficiently describing the spin chain eigenstates is much more difficult and has been the subject of active investigation over many years. A renewed interest in this problem stems from the appearance of higher rank (super-)spin chains in the context of computing correlation functions in $\cN=4$ supersymmetric Yang-Mills theory with the use of integrability (see e.g. \cite{Escobedo:2010xs,Gromov:2012vu,Vieira:2013wya,Caetano:2014gwa,Basso:2015zoa} and the review \cite{Beisert:2010jr}). Only for the simplest spin chains, which are based on the $SU(2)$-invariant R-matrix, can the states be obtained in a direct and compact way. In this case the algebraic Bethe ansatz approach allows one to build the eigenstates by acting on a reference state with a ``creation operator'' $B(u)$.
We found that similar operators $B^{\rm good}(u)$ can be constructed for any $SU(N)$ spin chain,
so that the eigenstates are given by
\beq
\label{B0}
	\ket{\Psi} = B^{\rm good}(u_1)B^{\rm good}(u_2)\dots B^{\rm good}(u_M)\ket{0}
\eeq
where $u_k$ are the Bethe roots. Furthermore, in constrast with the standard $SU(2)$
construction for $B(u)$, our $B^{\rm good}(u)$ is diagonalizable and
is suitable for immediate application of the separation of variables (SoV) approach as we describe below.
We provide an explicit expression for this operator as a polynomial in the monodromy matrix entries.

Our formula \eq{B0} provides a concise and compact alternative
to all the existing techniques for construction
of the eigenstates.
The most transparent method to obtain the states available in the literature is the \textit{nested} algebraic Bethe ansatz approach, in which the eigenvectors are built  recursively based on the solution of a lower rank spin chain \cite{SutherlandN1,Kulish:1983rd,Belliard:2008di}. Other known constructions include an explicit representation via sums over partitions of Bethe roots \cite{Belliard:2012sn}, as well as the rather sophisticated trace formulas of \cite{Tarasov1994} and the Drinfeld current construction \cite{Currents,BelPakR10} (see also \cite{Pakuliak:2014xqa,Belliard:2013mza,Hao:2016dyk,Albert:2000ne}).
These methods have been explored in-depth and have facilitated numerous calculations of various observables such as form factors and scalar products \cite{BelPakR10,Slavnov:2015qoa,Pakuliak:2015qga,Pakuliak:2014ela,Pakuliak:2015esa,	Belliard:2012is,Belliard:2012pr,Belliard:2012av,Pakuliak:2013zla,Pakuliak:2015fma,Pakuliak:2014zja}. Nevertheless in all of these approaches the expression for the states has a rather involved structure as well as being hard to implement computationally in many situations.\footnote{In particular, the result usually involves sums over partitions of Bethe roots or sums over states of an auxiliary spin chain, leading to exponentially many terms for highly excited states.}
In contrast to the nested Bethe ansatz, our construction involves no recursion and is also free from many of the complications inherent in other methods.


Although in the $SU(N)$ Bethe ansatz there are several types of Bethe roots, in the examples we considered it is just the momentum-carrying Bethe roots which should be plugged into the $B^{\rm good}$ operators in \eq{B0} to construct the states. The auxiliary Bethe roots enter the construction only indirectly, via the usual Bethe equations which determine the positions of the momentum-carrying roots\footnote{In fact the Bethe ansatz equations are not nessesarily the best way of finding Bethe roots.
In many situations it is much more convenient to solve directly the $N$-th order Baxter equation involving no auxiliary Bethe roots.
This method gives directly the Baxter polynomial containing the momentum-carrying roots only.}. Thus our construction is also free from the ambiguity associated with different possible
choices of one set of Bethe ansatz equations out of the
$N!$ equivalent possibilities (corresponding to different paths in the Hasse diagram).



Our construction is closely related to the Separation of Variables (SoV) approach.
At the classical level, SoV is a fundamental method of solving completely integrable
models which reduces the dynamics to a set of decoupled 1-particle problems.
While implementation of the SoV at the quantum level is more
subtle (in particular due to operator ordering issues), its
efficiency has also been demonstrated in many settings.
Broadly
speaking, the idea of the SoV method is to find  variables in
which the wavefunction of the system factorizes into 1-particle blocks.
The SoV in application to quantum integrable spin chains was pioneered
by Sklyanin in
\cite{Sklyanin:1991ss,Sklyanin:1995bm}.
Our proposal for the form of the operator $B^{\rm good}(u)$
is based on the Sklyanin's approach.





To ensure non-degeneracy of the construction we consider spin chains with a generic diagonal twist and generic inhomogeneities. Another important
element of our proposal is that one should add an extra similarity transformation in the auxiliary space in order to construct the eigenstates of the original spin chain.

Given the highly compact form of our representation for the states, we hope that our results should be useful in various contexts and  may also help to approach the longstanding problem of generalizing Slavnov's determinant expression for the scalar product beyond the $SU(2)$-type models. Exploring potential applications for spin chains with $PSU(2,2|4)$ symmetry would be especially interesting in view of their relation with the $\cN=4$ supersymmetric Yang-Mills theory.


Although the construction itself has a rather simple form, we found it is complicated to
prove it in general. Instead, we check the conjecture extensively on numerous examples and numerically, and
prove it in several particular cases. In this paper we focus on the canonical example of closed spin chains
with rational R-matrix and a fundamental representation of $SU(N)$ at each site\footnote{The usual R-matrix on which the spin chain is based is of course $GL(N)$-invariant, but we refer to the spin chain as having $SU(N)$ symmetry to emphasize that in the examples we consider we have a finite-dimensional representation at each site.}. It would be interesting to
find a general
proof algebraically.

This paper is organized as follows. In section 2 we introduce the relevant notation and the key definitions related to algebraic Bethe ansatz. In section 3 we discuss the $SU(2)$ case in detail and prove some novel nontrivial aspects of the construction involving our operator $B^{\rm good}$.
In section 4 we describe the SoV in the $SU(3)$ case and present our conjecture for the new construction of eigenstates in this setting. In section 5 we propose an extension of the SoV approach beyond the well-studied $SU(2)$ and $SU(3)$ cases, and describe our proposal which should provide eigenstates in any $SU(N)$ spin chain. Along the way we discuss many tests of the conjecture. In section 6 we present conclusions and outline future directions. Lastly, appendices contain some more technical details.

\section{Notation and basic definitions}

As the results we present in this paper are based on the algebraic Bethe ansatz framework, in this section we will describe its basic components for $SU(N)$ spin chains, and also introduce relevant notation.

In the algebraic Bethe ansatz approach the key object defining the integrable model is the R-matrix. The $SU(N)$ spin chains we study are based on the rational R-matrix which acts in $\mC^N \otimes \mC^N$ and has the form
\beq
	R_{12}(u)=(u-i/2) + iP_{12}
\eeq
where $P$ is the permutation operator and $u$ is the spectral parameter. We will concentrate on the case of spin chains with a fundamental representation of $SU(N)$ at each site. As usual we introduce an auxiliary space $\mC^N$ supplementing the physical Hilbert space $\cH$ which is a tensor product of $L$ copies of $\mC^N$,
\beq
	\cH=\mC^N\otimes\mC^N\otimes\dots \otimes\mC^N\ ,
\eeq
and construct the monodromy matrix as the product of R-matrices,
\beq
	T(u)=R_{01}(u-\theta_1)R_{02}(u-\theta_2)\dots R_{0L}(u-\theta_L) g\ ,
\eeq
where the R-matrix $R_{ak}$ acts in the tensor product of the auxiliary space $V_a$ and the $k$-th copy of $\mC^N$ inside the Hilbert space. The complex parameters $\theta_k$ are known as inhomogeneities, while $g$ is the twist matrix acting in the auxiliary space which we take to be diagonal,
\beq
	g={\rm diag}\(\lambda_1,\lambda_2,\dots,\lambda_N\)\ .
\eeq
The complex twists $\lambda_i$ and inhomogenieties $\theta_k$ serve as important regulators in our construction, and we assume they are all distinct and in generic position.

It is convenient to understand $T(u)$ as an $N\times N$ matrix,
\beq
\label{Tmat}
	T(u)=\begin{pmatrix}
		T_{11}(u)&\dots&T_{1N}(u)\\
		\vdots&\ddots&\vdots\\
		T_{N1}(u)&\dots&T_{NN}(u)
	\end{pmatrix}\ ,
\eeq
whose elements $T_{ij}(u)$ are operators acting only in the Hilbert space of the spin chain. These operators $T_{ij}$ satisfy nontrivial commutation relations that can be deduced from the RTT relation, which reads
\beq
\label{RTT}
	R_{ab}(u-v){T_a(u)} {T_b(v)}=
	{T_b(v)}{T_a(u)} R_{ab}(u-v)\ ,
\eeq
where we introduce two auxiliary spaces $V_a,V_b\simeq \mC^N$ so the R-matrix $R_{ab}$ acts in their tensor product, while ${T_k(u)}$ acts in $V_k\otimes\cH$.

The trace of $T(u)$ over the auxiliary space can be written as
\beq
	t_1(u)={\rm tr}_0 T(u)=\sum_{n=1}^N T_{nn}(u)\ , \
\eeq
and is an operator in the physical space $\cH$ known as the transfer matrix. It commutes with itself at different values of the spectral parameter,
\beq
	\[t_1(u),t_1(v)\]=0,\
\eeq
so that coefficients of the expansion of the transfer matrix in powers of $u$ form a commutative family of operators, a particular combination of which is in fact the Hamiltonian of the spin chain. These operators can therefore be diagonalized simultaneously. They also have a particularly simple eigenvector given by
\beq
\label{def0}
	\ket{0}=\begin{pmatrix}1\\0\\ \vdots \\ 0\end{pmatrix}\otimes\begin{pmatrix}1\\0\\ \vdots \\ 0\end{pmatrix}
	\otimes\dots \otimes \begin{pmatrix}1\\0\\ \vdots \\ 0\end{pmatrix}\ ,
\eeq
which is used as a reference/vacuum state in the algebraic Bethe ansatz approach we will discuss below. Let us finally mention that we will often use brief notation for shifts in the spectral parameter, namely
\beq\label{defpm}
	f^\pm\equiv f(u\pm i/2)\ , \ f^{[+a]}\equiv f(u+ia/2)\ .
\eeq

The main problem we focus on in this paper is constructing the common eigenvectors of the commuting operators $t_1(u)$, which are automatically also eigenstates of the spin chain Hamiltonian. In the next section we describe how this problem is solved for the $SU(2)$ spin chains.

\section{Eigenstates and SoV for the $SU(2)$ spin chain}

In this section we describe in detail the well-known construction of eigenstates and separated variables for the simplest $SU(2)$ spin chains. This case also illustrates several key elements of our approach which we will use later for higher rank spin chains.

\subsection{Algebraic Bethe ansatz and SoV in the $SU(2)$ case}

For $SU(2)$ spin chains the monodromy matrix $T(u)$ from \eq{Tmat} is a $2\times 2$ matrix whose entries act on the physical Hilbert space, and we denote them as
\beq
\label{TABCD}
	T(u)=\begin{pmatrix}
		A(u)&B(u)\\ C(u)& D(u)
	\end{pmatrix}\ .
\eeq
The transfer matrix whose eigenstates and eigenvalues we would like to obtain is then given by $A(u)+D(u)$, while the eigenstates can be built using $B(u)$ as
\beq
\label{B0su2}
	\ket{\Psi}=B(u_1)\dots B(u_M)\ket{0}\ .
\eeq
Here $\ket{0}$ is the vacuum state,
\beq
	\ket{0}=\begin{pmatrix}1 \\ 0\end{pmatrix}\otimes\dots\otimes \begin{pmatrix}1 \\ 0\end{pmatrix}\ ,
\eeq
while the Bethe roots $u_j$ are determined by the Bethe equations
\beq
	\prod_{n=1}^L\frac{u_j-\theta_n+i/2}{u_j-\theta_n-i/2}=\frac{\lambda_2}{\lambda_1}
	\prod_{k\neq j}^M\frac{u_j-u_k+i}{u_j-u_k-i}\ .
\eeq
Let us also note that
\beq
\label{Bcomm}
	\[B(u),B(v)\]=0\ ,
\eeq
so the states are symmetric in the Bethe roots.

Remarkably, in addition to generating the eigenstates the operator $B(u)$ has another important property -- namely, it provides the basis of separated variables within Sklyanin's SoV framework \cite{Sklyanin:1995bm}. Let us first summarize in general the SoV construction in the $SU(2)$ case, and then discuss several important subtleties regarding its implementation for our spin chain.

In order to construct the full set of separated variables, one assumes that $B(u)$ is a polynomial of degree $L$. Making explicit its zeros, one can write
\beq
\label{Bpsu2}
	B(u)=B_0\prod_{n=1}^L (u-x_n)\ ,
\eeq
where $B_0$ and $x_n$ are operators which commute due to \eq{Bcomm}\footnote{Several subtleties in defining the individual operators $x_n$ as operator zeros of $B$ are discussed in \cite{Sklyanin:1995bm,Sklyanin:1991ss}.}. At least in simplest examples, the operator $B_0$ is proportional to the identity operator on the Hilbert space so we can treat $B_0$ as a number. Assuming further that one can construct a complete set of left eigenvectors of $B(u)$ which form a basis in the Hilbert space, and labelling them as
$\bra{\rx_1,\dots,\rx_L}$ to indicate the eigenvalues of all $x_n$, we find from \eq{B0su2}\footnote{As discussed in \cite{Sklyanin:1995bm}, algebraically one can deduce various important properties of these variables even without using the construction of eigenstates \eq{B0su2} (that construction in particular assumes the existence of a good reference state $\ket{0}$ which is not true for some models). }
\beq
	\bra{\rx_1,\dots,\rx_L}\Psi\rangle=\bra{\rx_1,\dots,\rx_L}0\rangle B_0 \prod_{j=1}^M\prod_{n=1}^L (u_j-\rx_n)\ .
\eeq
Then normalizing the eigenvectors so that $\bra{\rx_1,\dots,\rx_L}0\rangle B_0=1$
we finally see that
\beq
\label{facsu2}
	\bra{\rx_1,\dots,\rx_L}\Psi\rangle=\prod_{n=1}^L (-1)^MQ_1(\rx_n)\ ,
\eeq
where $Q_1(u)$ is the Baxter Q-function,
\beq
	Q_1(u)=\prod_{j=1}^M(u-u_j)\
\eeq
and the extra sign $(-1)^M$ comes from rearranging the product.
In other words, \eq{facsu2} means that the wavefunction in this basis is factorized, realizing the key goal of the SoV approach. This simple form of the wavefunction greatly facilitates the computation of various observables such as form factors or scalar products (see \cite{Kitanine:2016pvg, Kitanine:2015jna, Kitanine:2014swa,Kazama:2013rya,Niccoli:2012ci,Levy-Bencheton:2015mia,Niccoli:2014sfa} for some interesting recent examples).

We see that the factorization of the wavefunction in the eigenbasis of $B(u)$ is clear almost at once from the construction of eigenstates \eq{B0su2}. In the next section we will discuss details of the SoV approach in an explicit example.

\subsection{Implementation for the $SU(2)$ spin chain}

\label{sec:su2impl}

While the above construction of separated variables should apply in principle to a wide variety of integrable models based on the rational $SU(2)$ R-matrix, it is not completely trivial to implement it in practice for the spin chain we consider. The main problem is that the operator $B(u)$ is nilpotent and not diagonalizable\footnote{In fact it's intuitively clear that $B$ is nilpotent in this case, as one can use it repeatedly to create all the states starting from the vacuum $\ket{0}$ which is a 'highest weight' state, and eventually reaching the maximally excited 'lowest weight' state which is annihilated by $B$. }, even with generic twists $\lambda_j$ and inhomogeneities $\theta_k$. This degeneracy spoils the construction, since it is then not possible to construct an eigenbasis for the operators $x_n$. Moreover $B(u)$ is a polynomial of degree less than $L$, making it even more problematic to define $L$ nontrivial operators $x_n$.


To circumvent this problem it is convenient to introduce an extra similarity transformation in the auxiliary space. Namely, instead of $T(u)$ we will consider a new matrix\footnote{A similar transformation was considered independently in \cite{Jiang:2015lda}, and we would like to thank I.~Kostov and D.~Serban for
illuminating discussions related to this approach. Parts of the construction were also discussed in \cite{SklyaninSchr}.
}
\beq
\label{Tgdef}
	T^{\rm good}(u)=K^{-1} T(u) K\ ,
\eeq
where the $2\times 2$ constant matrix $K$ acts only in the auxiliary space. This extra twisting will also be important for the $SU(N)$ spin chains that we discuss below. Notice that this transformation leaves unchanged the trace of $T(u)$, i.e. the transfer matrix which we want to diagonalize. Moreover, due to $GL(2)$ invariance of the R-matrix the new monodromy matrix $T^{\rm good}(u)$ will satisfy the same RTT relations \eq{RTT} as $T(u)$, so commutation relations between the elements $T_{ij}^{\rm good}(u)$ will be the same as before. We will label the entries of $T^{\rm good}$ similarly to \eq{TABCD},
\beq
	T^{\rm good}(u)=\begin{pmatrix}A^{\rm good}(u)&B^{\rm good}(u)\\ C^{\rm good}(u)&D^{\rm good}(u)\end{pmatrix}\ .
\eeq

The key point is that now one can use $B^{\rm good}(u)$ to construct both the separated varaibles and the eigenstates of the orginal spin chain.
With a generic choice of the matrix $K$, the operator $B^{\rm good}(u)$ will be diagonalizable and also of degree $L$ as a polynomial in $u$, as needed. As the commutation relations are unchanged, we also have
\beq
		[B^{\rm good}(u),B^{\rm good}(v)]=0\ .
\eeq
As for the eigenstates, let us first focus on a simple example when $K$ is upper triangular,\footnote{Without loss of generality we can assume $\det K=1$ as the scalar part of $K$ drops out of \eq{Tgdef}.}
\beq
\label{Kab}
	K=\begin{pmatrix}a&b\\ 0&1/a\end{pmatrix}\
\eeq
(note that a particularly simple choice would be $a=b=1$). This already guarantees that if $a,b$ are nonzero then $B^{\rm good}(u)$ is of degree $L$ in $u$ as one can easily check. Moreover, the matrix $T^{\rm good}$ then reads
\beq
\label{Tg0}
	T^{\rm good}(u)=
	\begin{pmatrix}
	A(u)-abC(u)\ \ \ &\frac{b}{a}\(A(u)-D(u)\)+\frac{1}{a^2}B(u)-b^2C(u)\\
	a^2 C(u)& D(u)+abC(u)
	\end{pmatrix}\ ,
\eeq
and as the vacuum state $\ket{0}$ is annihilated by $C(u)$, we see that
\beq
\label{Ag0}
	A^{\rm good}(u)\ket{0}=A(u)\ket{0},\ \ \ D^{\rm good}(u)\ket{0}=D(u)\ket{0}\ \ .
\eeq
Let us recall that to prove that \eq{B0su2} are eigenstates of the transfer matrix it is enough to use the commutation relations between the $A,B,C,D$ operators, together with the fact that the vacuum is an eigenstate of $A(u)$ and $D(u)$ with eigenvalues of a prescribed form. As we discussed, $T^{\rm good}(u)$ satisfies the same RTT relations as $T(u)$, which in combination with relations \eq{Ag0} guarantees that $B^{\rm good}$ can be used instead of $B$ to build eigenstates as in \eq{B0su2}, i.e.\footnote{We use the $\propto$ sign to indicate that normalization of the states \eq{B0su2} and \eq{Bgsu2} may be different}.
\beq
\label{Bgsu2}
	\ket{\Psi}\propto B^{\rm good}(u_1)\dots B^{\rm good}(u_M)\ket{0}
\eeq

Quite surprisingly, the operator $B^{\rm good}$ turns out to generate the eigenstates even for a generic matrix $K$. The proof of this fact is considerably more involved, the main difficulty being that in this case $\ket{0}$ is no longer an eigenstate of $A^{\rm good}$ and $D^{\rm good}$. Nevertheless we found a proof and present it in Appendix \ref{app:proofsu2}.  The main idea is to apply the SoV approach and make use of the variables conjugated to the separated variables $x_n$. Our only assumption is that the spectrum of $x_n$ is $\theta_n\pm i/2$ as we discuss below\footnote{We are grateful to E.~Sklyanin for pointing out to us, after this paper was finished, that a different proof of essentially the same statement for the $SU(2)$ case has appeared before in \cite{SklyaninSchr}.}.


Having constructed the operator $B^{\rm good}$ we can study in detail the implementation of the SoV program. As expected, we observe that the eigenvalues of each $x_n$ are $\theta_n\pm i/2$ (see e.g. \cite{Sklyanin:1995bm})\footnote{Strictly speaking there is an ambiguity in the definition of individual $x_n$ operators, as the coefficients of the polynomial $B(u)$ involve only their symmetric combinations. It is natural to define them in such a way that each $x_n$ is associated with a particular site of the chain and has eigenvalues $\theta_n\pm i/2$ determined by the inhomogeneity at that site. This choice fixes the operators $x_n$ uniquely.}. Then we can label the common eigenbasis of all $x_n$ by the choice of signs in the eigenvalues, e.g. the state $\bra{++\dots+}$ corresponds to all signs chosen as $+$. As we need to choose $L$ signs, we get a complete basis of $2^L$ states. As an example, from \eq{facsu2} we get
\beqa
	\bra{+++\dots +}\Psi\rangle &=&(-1)^{LM}
	Q_1\(\theta_1+\frac{i}{2}\)Q_1\(\theta_2+\frac{i}{2}\)Q_1\(\theta_3+\frac{i}{2}\)
	\dots Q_1\(\theta_L+\frac{i}{2}\)\ ,\
	\\ \nn
	\bra{+-+\dots +}\Psi\rangle &=&(-1)^{LM}
	Q_1\(\theta_1+\frac{i}{2}\)Q_1\(\theta_2-\frac{i}{2}\)Q_1\(\theta_3+\frac{i}{2}\)
	\dots Q_1\(\theta_L+\frac{i}{2}\)\ ,\
\eeqa
and so on.

A curious feature of the construction is that one can also build eigenstates with the same operator $B^{\rm good}$ but using a \textit{dual} set of Bethe roots and acting on a different reference state. This property is directly related to the fact that one can use an arbitrary matrix $K$ when building the operator $B^{\rm good}$ via \eq{Tgdef}. As this feature will also hold for the $SU(N)$ chains we describe below, let us discuss it in detail here for the $SU(2)$ case. Recall that in the usual $SU(2)$ algebraic Bethe ansatz one can alternatively build the states by starting with an alternative reference state
\beq
\label{su20p}
	\ket{0'}=\begin{pmatrix}0 \\ 1\end{pmatrix}\otimes\dots\otimes \begin{pmatrix}0 \\ 1\end{pmatrix}\ ,
\eeq
and acting on it with $C(u)$ rather than $B(u)$,
\beq
\label{sC}
	\ket{\Psi'}=C(v_1)\dots C(v_{L-M})\ket{0'}
\eeq
with operators $C$ evaluated at the dual Bethe roots $v_i$ which are zeros of the polynomial $Q_2$ defined by
\beq
	\lambda_2Q_1\(u+\frac{i}{2}\)Q_2\(u-\frac{i}{2}\)
	-{\lambda_1}Q_1\(u-\frac{i}{2}\)Q_2\(u+\frac{i}{2}\)=(\lambda_2-\lambda_1)\prod_{n=1}^L(u-\theta_n)\ .
\eeq
In our construction there is no need to switch between the operators $B$ and $C$, instead one can use the \textit{same}  operator $B^{\rm good}$ and act on the different reference state using the dual set of Bethe roots,
\beq
\label{Bgdsu2}
	\ket{\Psi'}\propto B^{\rm good}(v_1)\dots B^{\rm good}(v_{L-M})\ket{0'}\ .
\eeq
This can be proven using the same arguments as for the construction with usual Bethe roots and a generic matrix $K$ which we discissed above (see Appendix \ref{app:proofsu2}).


\subsection{The scalar product in the SoV representation}

\label{norm1}
In this section we discuss briefly the scalar products of two 'off-shell' states.
For convenience we pick a one-parametric family of $B^{good}$ parameterized
with an upper triangular matrix $K$ in \eq{Tgdef}
of the form
\beq
	K=
	\begin{pmatrix}
	\frac{1}{\sqrt{\alpha}}\ \ &-\frac{1}{\sqrt{\alpha}}
	\\
	0&\sqrt{\alpha}
	\end{pmatrix}\ .
\eeq
Here $\alpha$ is an unfixed parameter which we retain for convenience. To make the results more compact we also choose to multiply $B^{\rm good}$ by an overall  factor $1/(\lambda_2-\lambda_1)$, so that explicitly we have
\beq
\label{Balp}
	B_{\alpha}^{\rm good}=\frac{1}{\lambda_2-\lambda_1}\(\alpha B-1/\alpha\; C-A+D \) \ .
\eeq
In this subsection we also assume that the twist matrix is an $SU(2)$ element, i.e.
\beq
	\lambda_1=1/\lambda_2=1/\lambda_1^* \ .
\eeq
We can define two Bethe vectors by:
\beq
|\psi\rangle=B_{\alpha}^{\rm good}(u_1)\dots B_{\alpha}^{\rm good}(u_N)\ket{0}\;\;,\;\;
|\phi\rangle=B_{-\frac{1}{\bar\alpha}}^{\rm good}(v_{1})\dots B_{-\frac{1}{\bar\alpha}}^{\rm good}(v_{L-N})\ket{0'}\;.
\eeq
Let us note that in the second equation here we use $B^{\rm good}$ evaluated with $\alpha$ in \eq{Balp} replaced by $-1/{\bar \alpha}$. If the Bethe roots $u_i$ satify the corresponding Bethe equations (and dual Bethe equations for $v's$) we get 
two eigenstates of the Hamiltonian. We can define off-shell
Bethe states when this requirement is relaxed. We will see that it is not
required for the computation of the scalar products of these two vectors. Using that $(B_{\alpha}^{\rm good}(u))^\dagger=B_{-1/\bar\alpha}^{\rm good}(\bar u)$ (i.e. Hermitian conjugation corresponds to replacing $\alpha$ by $-1/{\bar \alpha}$) we find that
\beq
\label{zbbz}
\langle	\phi|\psi\rangle=\bra{0'}B_\alpha^{\rm good}(u_1)\dots B_\alpha^{\rm good}(u_M)\ket{0}\
\eeq
where $u_i$ are the original Bethe roots for $i=1..N$ and $u_i=v_{i-N}$ for $i>N$.

The main idea is to write a representation of the identity operator in terms of left and right eigenvectors of $B^{\rm good}$ (in the remainder of this section we omit the subscript $\alpha$). We already considered the left eigenvectors $\bra{\rx_1,\dots,\rx_L}$ which we will denote for brevity as $\bra{\rm x}$, so we have\footnote{With the definition \eq{Balp} that we are using, the leading coefficient of $B^{\rm good}$ as a polynomial in $u$ is $1$.}
\beq
	\bra{\rx}B^{\rm good}(u)=\bra{\rx}\prod_{n=1}^L(u-\rx_n)\ .
\eeq
One can easily see (at least for the first few $L$'s) that the right eigenvalues of $B^{\rm good}$ are the same as the left ones. We will denote the right eigenstates as $\ket{\rx}$, although they are not Hermitian conjugates of the left eigenvectors $\bra{\rx}$. Then we have
\beq
	B^{\rm good}(u)\ket{\rx}=\prod_{n=1}^L(u-\rx_n)\ket{\rx}\ .
\eeq


While the left eigenstates are normalized such that
\beq
	\ \bra{\rx}0\rangle=1\ ,
\eeq
it is convenient to normalize the right states using the dual vacuum $\ket{0'}$ so that
\beq
	\bra{0'}\rx\rangle=1\ .
\eeq
Then the key observation is that the scalar product of the left and the right states is simple and reads\footnote{We have checked this explicitly for the first few $L$'s with high numerical precision, but a complete proof should also not be too difficult.}
\beq
	\bra{\rx}{\rm y}\rangle=\frac{1}{\mu(\rx)}\delta_{\rx,{\rm y}}
\eeq
with the measure $\mu$ given by\footnote{Similar expressions for the measure already appeared in \cite{Kazama:2013rya,Jiang:2015lda}. }
\beq
\label{mux}
	\mu(\rx)=\(\frac{\alpha \lambda_2}{\lambda_1-\lambda_2}\)^L
	\prod_{n=1}^L\(\frac{\rx_n-\theta_n}{i/2}\)\;
 \frac{\prod\limits_{m<n}^L(\bar\rx_m-\bar\rx_n)}{\prod\limits_{m<n}^L(\theta_m-\theta_n)}\ ,
\eeq
where we denoted
\beq
	\bar{\rx}_n=
	\begin{cases}
	\theta_n-i/2,& \text{for } \rx_n=\theta_n+i/2\\
	\theta_n+i/2,& \text{for } \rx_n=\theta_n-i/2
	\end{cases}\ .
\eeq
This means that we can write a resolution of the identity as
\beq
	{\bf 1}=\sum\limits_{\rx_n=\theta_n\pm i/2}\ket{\rx}\bra{\rx}\;\mu(\rx)\ .
\eeq
Inserting this representation of the identity operator immediately in front of the vacuum $\ket{0}$ in the scalar product \eq{zbbz}, we get at once
\beqa
	\bra{0'}B^{\rm good}(u_1)\dots B^{\rm good}(u_L)\ket{0}&=&
	\sum_{\rx_n=\theta_n\pm i/2}\mu(\rx)\prod_{m=1}^M\prod_{n=1}^L(u_m-\rx_n)\ .
\eeqa
The left-hand side of this equation is actually zero for $M<L$, but it still holds for any $M$. Thus as expected we can write this scalar product as a sum over all values of separated variables, with a simple measure \eq{mux}.

In the appendix \ref{app:norm} we also link the usual Gaudin formula for the norm of a transfer matrix eigenstate with the SoV approach and also with the variation of the Q-functions with respect to the inhomogeneity parameters $\theta_i$. In particular we rewrite the Gaudin formula in terms of Q-functions only, rather than in terms of individual Bethe roots (the result is given by \eq{N1app}). Similar results already appeared in \cite{Kitanine:2016pvg,Kitanine:2015jna,Kazama:2013rya}, but our derivation is different and utilizes several tricks which may be useful in other contexts.

\section{Eigenstates for $SU(3)$ spin chains from SoV}

In this section we review how the SoV approach works for $SU(3)$ spin chains. We then discuss in detail our main conjecture for this case, which states that spin chain eigenstates can be generated using the same operator that provides separated variables.

\subsection{The $B$ operator}

The operator $B(u)$ which should provide separated variables for quantum $SU(3)$-type models was constructed by Sklyanin in \cite{Sklyanin:1992sm} (following the construction in the classical case \cite{Sklyanin:1992eu}). It is built from entries of the monodromy matrix which in this case is a $3\times 3$ matrix
\beq
	T(u)=\begin{pmatrix}
		T_{11}(u)&T_{12}(u)&T_{13}(u)\\
		T_{21}(u)&T_{22}(u)&T_{23}(u)\\
		T_{31}(u)&T_{32}(u)&T_{33}(u)
	\end{pmatrix}
\eeq
whose entries act on the physical Hilbert space. To build the operator $B$ it is convenient to introduce the $2\times 2$ quantum minors of this matrix as
\beq
\label{min22}
	T_{j_1j_2|k_1k_2}(u)=T_{j_1k_1}(u)T_{j_2k_2}(u+i)-T_{j_2k_1}(u)T_{j_1k_2}(u+i)\ .
\eeq
In the $SU(3)$ case it is natural to assemble these minors into a new matrix $U$ as
\beq
	U_{jk}(u)=(-1)^{j+k}T_{\bar{j}|\bar{k}}(u)
\eeq
where $\bar{j}=\{1,2,3\} \backslash \{j\}$. In fact $U$ is also the monodromy matrix constructed for the case when the auxiliary space carries the antisymmetric representation of $SU(3)$ rather than the fundamental representation (in particular, $U$ satisfies the same RTT relations \eq{RTT} as the monodromy matrix $T$). With these definitions, the operator $B(u)$ of \cite{Sklyanin:1992sm} is given by
\beq
\label{BU}
	B(u)=T_{23}(u)U_{31}(u-i)-T_{13}(u)U_{32}(u-i)\ ,
\eeq
or explicitly
\beqa
\label{B3}
	B(u)&=&T_{23}(u)T_{12}(u-i)T_{23}(u)-T_{23}(u)T_{22}(u-i)T_{13}(u) \\ \nn
	&+&
	T_{13}(u)T_{11}(u-i)T_{23}(u)-T_{13}(u)T_{21}(u-i)T_{13}(u)
	\ .
\eeqa
As an indication of the complexity of the $SU(3)$ models, let us note that $B(u)$ is a polynomial of the 3rd degree in the monodromy matrix entries, while in the $SU(2)$ case it was simply the element $T_{12}$.

From the commutation relations between elements of $T(u)$ it follows that, remarkably,
\beq
	[B(u),B(v)]=0
\eeq
so that similarly to the $SU(2)$ case illustrated in \eq{Bpsu2} one can introduce new commuting operators as zeros of $B(u)$. Importantly, in the $SU(3)$ case $B(u)$ is expected to be a polynomial of degree $3L$ in $u$ as discussed in \cite{Sklyanin:1992sm}, so we have three zeros for each site of the chain. We will see below that based on the eigenvalue spectrum of $B(u)$ it is natural to split the new operators into $L$ groups with three operators in each group, so we can write
\beq
\label{Bx3}
	B(u)=B_0\prod_{n=1}^{L}\prod_{a=1}^3(u-x_{n,a})\ ,\
\eeq
where the index $n$ labels the sites, while $a$ enumerates the three $x$ operators associated with this site.
The leading coefficient $B_0$ also commutes with all $x_n$.

\subsection{Application to spin chains and construction of eigenstates}

Let us discuss how to apply the above construction to our main example, the spin chain with a fundamental representation at each of the $L$ sites. Calculating $B(u)$ explicitly for low values of $L$ we immediately see that as in the $SU(2)$ case it is nilpotent and not diagonalizable.
In fact it annihilates almost the entire Hilbert space (for instance, when $L=1$ all its matrix elements are simply zero). This is true even with generic twists $\lambda_i$ and inhomogeneities $\theta_n$. To remedy this problem we again apply a similarity transformation to the monodromy matrix as we described in \eq{Tgdef} for the $SU(2)$ case, so we introduce
\beq
\label{Tgd3}
	T^{\rm good}(u)=K^{-1}T(u)K
\eeq
where $K$ is a constant matrix for which a particularly simple choice is
\beq
\label{K3}
	K=\begin{pmatrix}
	1&1&1\\
	0&1&1\\
	0&0&1
	\end{pmatrix}\ .
\eeq
As we will discuss below, we observed that even for a generic $K$ all the key features of our construction are preserved.

Now we build a new operator $B^{\rm good}(u)$ by using the expression \eq{B3} for $B(u)$ in which we replace the elements $T_{ij}(u)$ by $T^{\rm good}_{ij}(u)$. All algebraic properties of the original $B(u)$ will be inherited by $B^{\rm good}$ as the commutation relations between elements of $T$ are preserved by the transformation \eq{Tgd3}, in particular we have
\beq
	[B^{\rm good}(u),B^{\rm good}(v)]=0\ .
\eeq
At the same time, computing $B^{\rm good}$ explicitly for various $L$'s we observe that it now can be diagonalized and moreover is a polynomial of degree $3L$ as required, with a nonvanishing leading coefficient given by
\beq
	B_0=(\lambda_1-\lambda_2)(\lambda_1-\lambda_3)(\lambda_2-\lambda_3)\ .
\eeq
Then the separated variables $x_n$ will be defined as zeros of $B^{\rm good}$,
\beq
\label{Bgx3}
	B^{\rm good}(u)=B_0\prod_{n=1}^{L}\prod_{a=1}^3(u-x_{n,a}) \ .
\eeq

Our key conjecture is that in addition to providing separated variables, the operator $B^{\rm good}(u)$ also generates the eigenstates of the transfer matrix, which can be built exactly as in the $SU(2)$ case, namely
\beq
\label{B0su3}
	\ket{\Psi}=B^{\rm good}(u_1)\dots B^{\rm good}(u_M)\ket{0}\ ,
\eeq
where $\ket{0}$ is the usual reference state
\beq
\label{su30}
	\ket{0}=\begin{pmatrix}1\\0\\0\end{pmatrix} \otimes \dots \otimes \begin{pmatrix}1\\0\\0\end{pmatrix}\ .
\eeq
In this formula the parameters $u_k$ are the momentum-carrying Bethe roots. They are determined by the usual Bethe equations which describe the spectrum of the transfer matrix eigenvalues and are well-known from the nested Bethe ansatz approach. These equations read
\beqa
	\prod_{n=1}^L\frac{u_j-\theta_n+i/2}{u_j-\theta_n-i/2}&=&\frac{\lambda_2}{\lambda_1}
	\prod_{k\neq j}^M\frac{u_j-u_k+i}{u_j-u_k-i}\prod_{k=1}^R\frac{u_j-v_k-i/2}{u_j-v_k+i/2}\ ,
	\\ \nn
	\prod_{n=1}^M\frac{v_j-u_n+i/2}{v_j-u_n-i/2}&=&\frac{\lambda_3}{\lambda_2}
	\prod_{k\neq j}^R\frac{v_j-v_k+i}{v_j-v_k-i}\ .
\eeqa
The parameters $v_k$ are known as the auxiliary Bethe roots and they do not enter directly our construction of the states. Their only role is that they appear in the equations above which fix the positions\footnote{Let us also note that for a fixed number of magnons $M\leq L$, there can be several states, parameterized in particular by the number of auxiliary roots which should satisfy $R\leq M$.}  of the momentum-carrying roots $u_n$.

In summary, we expect that to build the full set of states one should go through all solutions of the Bethe equations, and for each of them plug the corresponding set of $u_j$ into the formula \eq{B0su3}.

For comparison let us briefly recall the well-known construction of the states via the nested algebraic Bethe ansatz. In this approach they are given by
\beq
\label{su3ba}
	\ket{\Psi}=\sum_{a_i=2,3}F^{a_1a_2\dots a_M}T_{1a_1}(u_1)T_{1a_2}(u_2)\dots T_{1a_M}(u_M)\ket{0}
\eeq
where $F^{a_1a_2\dots a_M}$ is the wavefunction of an auxiliary $SU(2)$ spin chain with $M$ sites in which $u_j$ are the inhomogenieties. The auxiliary Bethe roots $v_k$ should be understood as the rapidities of magnons propagating on this spin chain. Notice that two different operators act on the vacuum here ($T_{12}$ and $T_{13}$), which moreover do not commute with each other. In contrast, our conjecture \eq{B0su3} features only one operator $B^{\rm good}(u)$ which also commutes with itself at different values of $u$. Furthermore, \eq{su3ba} is a sum of $2^M$ nontrivial terms, with coefficients $F^{a_1a_2\dots a_M}$ that are quite complicated and are determined recursively from a lower rank spin chain. Our formula instead is much more compact and has the same form as for the simplest $SU(2)$ spin chain.

%

We have checked our conjecture in a multitude of cases, leaving little doubt in its validity. At the same time, obtaining a complete analytic proof is likely to be quite nontrivial. In the next section we describe various tests as well as presenting a proof for some special cases.

\subsection{Tests of the proposal for eigenstates}

We have numerically checked our conjecture extensively using Mathematica for values of $L$ up to about $L\sim 5$.  We considered many states, with various numbers of roots of each type, and found that our proposal \eq{B0su3} works perfectly. In order to easily solve the Bethe equations, we typically considered a configuration with generic real twists $\lambda_i\sim 100$ and large real inhomogeneities $\theta_i \sim 100$ which are also well separated from each other, i.e. $|\theta_i-\theta_j| \sim 100$. In this regime the values of all the Bethe roots are close to the inhomogeneities, so having these good starting points it is simple to find very accurate numerical solutions of the Bethe equations. Then we checked that our proposal provides eigenstates with very high precision (typically $50-100$ digits). In the $SU(N)$ cases which will be discussed below in this paper we followed a similar numerical approach.

In addition, we have proven the conjecture analytically for \textit{arbitrary} $L$ for states with $1$ and $2$ momentum-carrying roots. The proof is done for the case when the matrix $K$ which we use to build $B^{\rm good}$ via \eq{Tgd3} is taken to be the upper triangular matrix \eq{K3}. The key point is that in this case $\ket{0}$ is an eigenstate for most of the elements $T_{ij}$ (see e.g. \cite{Belliard:2012sn}), namely
\beqa
\label{T110}
	T_{11}(u)\ket{0}&=&\lambda_1 Q_\theta^+(u)\ket{0},\ \ \
	T_{22}(u)\ket{0}=\lambda_2 Q_\theta^-(u)\ket{0},\ \\
	\label{T330}
	T_{33}(u)\ket{0}&=&\lambda_3 Q_\theta^-(u)\ket{0}\ ,
\eeqa
and
\beq
\label{T210}
	T_{21}(u)\ket{0}=T_{31}(u)\ket{0}=T_{32}(u)\ket{0}=T_{23}(u)\ket{0}=0\ .
\eeq
Here we denoted
\beq
	Q_\theta=\prod_{n=1}^L(u-\theta_n)\
\eeq
and we recall the notation from \eq{defpm},
\beq
	f^\pm\equiv f(u\pm i/2)\ , \ f^{[+a]}\equiv f(u+ia/2)\ .
\eeq
The operator $B^{\rm good}$ which we use is constructed out of elements of $T^{\rm good}$, for which we have precisely the same relations \eq{T110}, \eq{T330}, \eq{T210} with the only exception being
\beq
	T_{23}^{\rm good}(u)\ket{0}=\(\lambda_2 Q_\theta^--\lambda_3 Q_\theta^-\)\ket{0}\ .
\eeq
Then our strategy is to act on the state \eq{B0su3} with the transfer matrix, and commute all the elements of $T^{\rm good}$ (except $T^{\rm good}_{12}$ and $T^{\rm good}_{13}$) to the right until they hit the vacuum $\ket{0}$ which is an eigenstate for them. To do this we use the commutation relations between elements of $T^{\rm good}$ which follow from the RTT relation \eq{RTT}. As a result the expression greatly simplifies, and one can also see that if the Bethe equations are satisfied several terms will cancel. As an example, for a state with one momentum-carrying root built as
\beq
\label{su31m}
	\ket{\Psi}=B^{\rm good}(u)\ket{0}
\eeq
we find
\beqa
\nn
	\(T^{\rm good}_{11}(w)+T^{\rm good}_{22}(w)+T^{\rm good}_{33}(w)\)B^{\rm good}(u)\ket{0}=&& \\
	\label{TgB}
	f_1T_{12}^{\rm good}(u)\ket{0}+f_2T_{13}^{\rm good}(u)\ket{0}+f_3T_{13}^{\rm good}(w)\ket{0}&&\ ,
\eeqa
where $f_i$ are some lengthy coefficients. We see that the last term here is unwanted, as it includes $T_{13}^{\rm good}(w)$ which clearly cannot appear in the expression for the state itself. However, its coefficient $f_3$ is rather simple,
\beq
	f_3=-\frac{i(\lambda_3-\lambda_2)Q_\theta^{[-3]}(u)}{w-u}
	\(\lambda_2 Q_\theta^-(u)-\lambda_1 Q_\theta^+(u)\)\(\lambda_3 Q_\theta^-(u)-\lambda_1 Q_\theta^+(u)\)\ .
\eeq
It's easy to see that on the solution of Bethe equations without auxiliary roots the next-to last factor in $f_3$ will be zero, while for the solution with one auxiliary root the last factor will vanish. This means that the unwanted term will disappear from \eq{TgB}. One can then check that the remaining part is proportional to the original state given by \eq{su31m} which is therefore indeed an eigenstate, as expected.

In a similar way we have proven that the construction works for states with 2 momentum-carrying roots
at any $L$. This calculation involves even more lengthy expressions and we do not give them here. In principle, the same approach should allow to prove the conjecture for any specified number of magnons. We hope however that a more algebraic proof can be found which would apply for any $L$ and any number of excitations at once.

\subsection{Construction with dual roots}

Another surprising feature of our construction is that one can use dual Bethe roots\footnote{See e.g. \cite{DualABA} for a pedagogical discussion of dualities in the Bethe ansatz.} as arguments of the operators $B^{\rm good}$ in \eq{B0su3}, provided they also act on the corresponding different reference state. This is in complete analogy with the $SU(2)$ case discussed above (see \eq{Bgdsu2}) where $B^{\rm good}$ can be used either with the usual roots or with the dual roots.  The main difference with the $SU(2)$ case is that now we have three natural reference states (defined as in \eq{su30} but with the $1$'s in the second row or in the third row), and accordingly to each eigenstate of the transfer matrix correspond three dual sets of momentum-carrying Bethe roots\footnote{In the Q-system language (see e.g. \cite{Gromov:2014caa} and references therein), these are the zeros of $Q_1(u),Q_2(u)$ and $Q_3(u)$.}. However, we observed that one and the same operator $B^{\rm good}$ allows to build the eigenstates starting from either reference state, as long as one uses the corresponding set of Bethe roots as arguments of $B$'s in \eq{B0su3}. We have checked this in many examples, but it would be interesting to find a rigorous proof.

This property is also directly linked with the fact that our construction works (as we observed in various examples) for a generic matrix $K$ in \eq{Tgd3}, as with arbitrary $K$ the reference state $\ket{0}$ is no longer distinguiushed. The possibility of taking $K$ generic also means that we have a whole family of operators $B^{\rm good}$ which all work well, and we found that this family is three-parametric in the $SU(3)$ case (a more detailed discussion is given in section \ref{sec:anyK}).

\subsection{The spectrum of separated variables}

To characterize the states in the SoV language, it is important to understand the eigenvalue spectrum of the operators $x_{n,a}$, which are the zeros of $B^{\rm good}(u)$ introduced in \eq{Bgx3},
\beq
	B^{\rm good}(u)=B_0\prod_{n=1}^{L}\prod_{a=1}^3(u-x_{n,a})\ .
\eeq
These operators define separated variables and we label their eigenstates according to their eigenvalues, as $\ket{\rx_{1,1},\dots,\rx_{L,3}}$. In direct analogy \eq{facsu2} for the $SU(2)$ case, our construction of states immediately guarantees that the wavefunction in the eigenbasis of $B^{\rm good}$ will take a factorized form,
\beq
	\bra{\rx_{1,1},\dots,\rx_{L,3}}\Psi\rangle=\prod_{n=1}^{L}\prod_{a=1}^3 (-1)^MQ_1(\rx_{n,a})
\eeq
where $Q_1(u)=\prod_{j=1}^M(u-u_j)$.

In the $SU(2)$ case there were as many $x_n$'s as sites, while here we have three variables $x_{n,a}$ for each site $n$ of the chain, since $B$ is a polynomial of degree $3L$. When $L=1$ the Hilbert space is $\mC^3$ and we found that the three left eigenvalues\footnote{It is the \textit{left} eigenvalues and eigenvectors of $B^{\rm good}$ which define the basis of separated variables, see the discussion for $SU(2)$ in the previous section.} of $B^{\rm good}$ read\footnote{The overall prefactor given here is the one obtained if one uses the upper triangular matrix $K$ from \eq{K3}.}
\beq
	(\lambda_1-\lambda_2)(\lambda_1-\lambda_3)(\lambda_2-\lambda_3)\(u-\(\theta_1+\frac{3i}{2}\)\)
	\(u-\(\theta_1-\frac{i}{2}\)\)^k\(u-\(\theta_1+\frac{i}{2}\)\)^{2-k}
\eeq
with $k=0,1,2$. We have verified this expression numerically with a high precision after setting $\lambda_i$ and $\theta_i$ to numerical values. There is an ambiguity in choosing which of the eigenvalues should be attributed to which of the $x$'s (as $B^{\rm good}$ is symmetric in all the $x_{n,a}$), but for the purpose of labelling the states we may choose e.g. the operator $x_{1,1}$ to act as a scalar equal to $\theta_1+\frac{3i}{2}$. Then there are three options for the remaining eigenvalues $\{\rx_{1,2},\rx_{1,3}\}$, namely
\beq
	\{\theta_1+ \frac{i}{2},\ \theta_1+ \frac{i}{2}\},\
	\{\theta_1+ \frac{i}{2},\ \theta_1- \frac{i}{2}\},\
	\{\theta_1- \frac{i}{2},\ \theta_1- \frac{i}{2}\}\ .
\eeq
We can label the corresponding three eigenstates according to the $\pm$ signs appearing in these eigenvalues, so we get three states
\beq
\label{pm1site}
	\ket{(++)},\ \ket{(+-)}, \ \ket{(--)}\
\eeq
which form a basis in the Hilbert space $\mC^3$. This is precisely the basis in which the wavefunction factorizes. Note that it is just the unordered set of eigenvalues which identifies the state uniquely, so we do not have an additional state $\ket{(-+)}$.

%


Furthermore, we observed that the pattern of eigenvalues for higher $L$ is obtained trivially from this one. Namely, $B^{\rm good}$ will contain a trivial overall scalar factor
\beq
	(\lambda_1-\lambda_2)(\lambda_1-\lambda_3)(\lambda_2-\lambda_3)
	\prod\limits_{n=1}^L\(u-\(\theta_n+\frac{3i}{2}\)\)
\eeq
so one can choose the $L$ operators $x_{k,1}$ with $k=1,\dots, L$ to act as scalars (with eigenvalues $\theta_k+\frac{3i}{2}$). The remaining nontrivial matrix part of $B^{\rm good}$ determines eigenvalues of the other
$x$'s, which are equal to
\beq
	\{\theta_1\pm \frac{i}{2}\ ,\theta_1\pm \frac{i}{2},\ \theta_2\pm \frac{i}{2},\ \theta_2\pm \frac{i}{2},
	\dots,\theta_L\pm \frac{i}{2},\ \theta_L\pm \frac{i}{2}\}\ .
\eeq
All combinations of signs are allowed here, and the unordered set of these eigenvalues labels the eigenstates of $B^{\rm good}$ giving $3^L$ possibilities which precisely corresponds to the dimension of the Hilbert space.

In other words, with each site $k$ we associate three operators $x_{k,1},x_{k,2},x_{k,3}$, one of which acts trivially on the whole space as $\theta_k+\frac{3i}{2}$ and the two others have eigenvalues $\theta_k\pm \frac{i}{2}$. At each site we can choose one of the three distinct combinations of signs in these eigenvalues, namely $++$, $+-$ and $--$.
To indicate all the nontrivial eigenvalues we can then label a state as e.g.
\beq
\label{eig3}
	\ket{(++)(+-)(+-)\dots}
\eeq
where we group in brackets the labels corresponding to the same site.



%
%
%

\section{Extension to the $SU(N)$ case}

In this section we demonstrate how to extend the compact construction of eigenstates to the $SU(N)$ spin chains. We first discuss the $SU(4)$ case and present the operator $B^{\rm good}(u)$, which we obtained by making an ansatz for it as a combination of quantum minors and then fixing the free parameters by several constraints. Then we extend the construction to the $SU(N)$ setting and discuss its features.

\subsection{The $B$ operator for $SU(4)$}

While the SoV program has been thoroughly studied for $SU(2)$-type models and to a lesser degree for the $SU(3)$ case (see e.g. \cite{Hao:2016dyk,Frahm:2015txa}), its extension to the case of higher rank groups presents a challenge. One of the key required ingredients is the operator $B(u)$ providing the separated variables. Let us motivate and present our conjecture for the form of this operator in the $SU(4)$ case. We will first discuss constructing the operator $B(u)$ itself and then, as before, we will apply an extra similarity transformation which removes degeneracies and provides the operator $B^{\rm good}(u)$ which generates the eigenstates of the spin chain.

The main inspiration comes from the form of $B$ given for $SU(3)$ by \eq{BU}, where it is written in terms of $2\times 2$ quantum minors of the monodromy matrix. For higher rank groups one can also construct $n \times n$ quantum minors which are known to satisfy various identities and frequently appear in the study of integrable models (see e.g \cite{MolevBook}). This suggests to make an ansatz for $B(u)$ using these building blocks. The $n\times n$ quantum minors are defined by a sum over permutations,
\beq\label{qmindef}
	T_{j_1,\dots,j_n|\; k_1,\dots,k_n}(u)=\sum_{\sigma\in S_n}(-1)^{{\rm sign}(\sigma)}
	T_{j_{\sigma(1)}k_1}(u)T_{j_{\sigma(2)}k_2}(u+i)\dots
	T_{j_{\sigma(n)}k_n}(u+(n-1)i)\ .
\eeq
In this notation $T_{i|j}(u)$ stands for the monodromy matrix element $T_{ij}(u)$ itself. The formula \eq{qmindef} is similar to a determinant but includes extra shifts in $u$. We can write the $SU(3)$ result explicitly as
\beq
\label{B3min}
	B(u)=T_{1|3}(u)T_{12|13}(u-i)+T_{2|3}(u)T_{12|23}(u-i)\ ,
\eeq
so for $SU(4)$ it would be natural to include also $3\times 3$ minors.

Another piece of information is the known form of $B(u)$ for \textit{classical} $SU(N)$ spin chains. It was found in \cite{Scott:1994dz,Ge95} (see also \cite{Adams:1992ej,Diener:1994xt} and the more recent work \cite{Falqui2007}) and reads \footnote{To be clear, in \eq{BclN} we denote by $T^n$ the product of $n$ matrices $T_{ij}$.}
\beq
\label{BclN}
	B=\sum_{i_1,\dots, i_{N-1}=1}^{N-1}
	\epsilon_{i_1\dots i_{N-1}} T_{i_1N} (T^2)_{i_2N}\dots (T^{N-1})_{i_{N-1}N}\ .
\eeq
We see that $B$ is a polynomial of degree $N(N-1)/2$ in the entries of the monodromy matrix, which here are all evaluated at the same value $u$ of the spectral parameter. However it is highly nontrivial to generalize this expression to the quantum case, as the classical limit corresponds to treating $T_{ij}$ as commuting elements so the operator ordering is lost. In addition, in this limit one removes all the shifts of the spectral parameter by multiples of $i$, even though e.g. in the quantum $SU(3)$ result \eq{B3min} the shifts play a key role.

One possible quantization of the classical result was proposed in \cite{Smirnov2001} in the context of models associated to $U_q(\widehat{sl}_N)$ symmetry, i.e. with a trigonometric R-matrix, although the implementation of this approach for concrete models was not discussed in that paper. While the result there is given in an implicit form, we independently derived an explicit expression for the $B$ operator which we will present shortly (and we also identified the role of $B$ for creating the transfer matrix eigenstates). It would be interesting though rather nontrivial to compare our results with \cite{Smirnov2001}, in particular we expect that an extra twist similar to $K$ we use to build $T^{\rm good}$ will be crucial for applying the construction of \cite{Smirnov2001} in explicit examples.

Motivated by the structure of \eq{BclN} and \eq{B3min}, we made the following ansatz for $B(u)$ in the $SU(4)$ case:
\beq
	B(u)=\sum s_{i_1i_2\dots i_6}T_{i_1|4}(u)T_{i_2i_3|i_4 4}(u+c_1)T_{123|i_5i_64}(u+c_2)
\eeq
where the sum runs over the values of indices
\beq
	i_1,i_2,\dots i_6=1,2,3\ \ \text{with}\ \ i_2<i_3,\ i_5<i_6\ .
\eeq
Here $s_{i_1\dots i_6}$ and $c_1,c_2$ are unfixed parameters. Note that due to the product of minors of increasing sizes in each term, $B$ is a polynomial of degree $1+2+3=6$ in the entries of $T_{ij}$, as required by the classical formula \eq{BclN}. We also ensured that $4$ appears among the column indices of each minor, to reflect the structure of the classical result.

To fix the unknowns in this ansatz we specialize to the case of the $SU(4)$ spin chain with fundamental representation at each site. Also, as before we construct from $B$ the operator $B^{\rm good}$, using instead of $T_{ij}$ the elements of $T^{\rm good}$ which is obtained by an upper triangular similarity transformation as in \eq{Tgd3}, \eq{K3}:
\beq
\label{Tg4}
	T^{\rm good}=K^{-1} T K,\ \ \ 	K=\begin{pmatrix}
	1&1&1&1\\
	0&1&1&1\\
	0&0&1&1\\
	0&0&0&1
	\end{pmatrix}\ .
\eeq
We use this concrete choice of $K$ when fixing the parameters in our ansatz, but we also found that once they are fixed the construction continues to work just as well with generic $K$.


Then in order to determine uniquely the unknown parameters $s_{i_1\dots i_6}$ and $c_{1,2}$ it is enough to impose that:
\begin{itemize}
	\item In the classical limit $B$ reduces to \eq{BclN}\footnote{In fact we impose that it should reduce to minus the classical result, this overall sign is irrelevant but gives a nicer expression for the quantum $B$ operator.}
	\item $[B^{\rm good}(u),B^{\rm good}(v)]=0$ for $L=1,2\ $
	\item For $L=1$ the operator $B^{\rm good}(u)$ generates all the three eigenstates with one momentum-carrying Bethe root by acting on the usual reference state $\ket{0}$, similarly to \eq{B0su3}
\end{itemize}
Remarkably, this fixes most of the coefficients $s_{i_1\dots i_6}$ to zero, while the rest are set to be equal to 1! Another nice feature is that the shifts $c_{1,2}$ are fixed to successive multiples of $i$, namely
\beq
	c_1=-i,\ c_2=-2i\ .
\eeq
Thus the expression for $B$ can be written simply as
\beq
\label{B4}
	B(u)=\sum_{j,\,k}T_{j|4}(u)T_{k|j4}(u-i)T_{123|k4}(u-2i)
\eeq
where $k=\{k_1,k_2\}$ and the sum runs over the values of indices $j,k_1,k_2$ from $1$ to $3$ with $k_1<k_2$.

We have checked this result extensively. In particular we find that the property
\beq
\label{Bcomm4}
	[B^{\rm good}(u),B^{\rm good}(v)]=0\ ,
\eeq
which we ensured for $L=1,2$, continues to hold for $L=3,4$. Most importantly, we found that eigenstates of the transfer matrix can be generated as before,
\beq
	\ket{\Psi}\propto B^{\rm good}(u_1)\dots B^{\rm good}(u_M)\ket{0}
\eeq
where $u_k$ are the momentum-carrying Bethe roots.

Thus we have extended our conjecture for construction of eigenstates to the $SU(4)$ case. We have numerically\footnote{See the beginning of section 4.3 for a brief summary of our numerical strategy.} checked it thoroughly for values of $L$ up to $L=4$ and various numbers of the Bethe roots of different types excited\footnote{Note that for $L=4$ we already have to deal with rather large matrices having $256\times 256=65536$ elements.}. In particular, for $L=2$ we verified numerically that the construction gives \textit{all} the 16 states, both for a generic matrix $K$ and for the upper triangular $K$ from \eq{Tg4}. We expect that the same operator $B$ should also provide the separated variables as its operator zeros, similarly to the $SU(2)$ and $SU(3)$ cases.

Let us also mention that another expression for the operator $B(u)$ which should provide separated variables for rational $SU(N)$ models was proposed in \cite{Chervov:2007bb}, motivated by considerations related to so-called Manin matrices. In the $SU(4)$ case we observed that, as expected, without the extra twist matrix $K$ that operator is nilpotent and cannot be diagonalized (at least for $L=1,2$). Using a nontrivial $K$ matrix as in \eq{Tg4} cures this problem, but we found that already for $L=2$ the resulting operator does not satisfy the  commutativity property \eq{Bcomm4} and therefore cannot be used to define separated variables\footnote{Accordingly, it also does not generate the two-magnon eigenstates of the transfer matrix already for $L=2$.}. The structure of our result is nevertheless rather similar to that in \cite{Chervov:2007bb}, and it would be interesting to look for possible connections.


\subsection{Generalization to any $SU(N)$}

Let us now present the conjecture for the operator $B$ generalizing the above results to any $SU(N)$ group. To see the structure let us write again the $SU(3)$ and $SU(4)$ results from \eq{B3min}, \eq{B4}, which read
\beqa
	B^{SU(3)}(u)=\sum_{j=1}^2&&T_{j|3}(u)T_{12|j3}(u-i)\ , \\ \nn
	B^{SU(4)}(u)=\sum\limits_{j,k}^3&& T_{j|4}(u)T_{k|j4}(u-i)T_{123|k4}(u-2i) \ .
\eeqa
Our conjecture for $SU(N)$ is
\beq
\label{BqN}
	B(u)=\sum_{j,k,\dots, p}T_{j|N}(u)T_{k|jN}(u-i)T_{l|kN}(u-2i)...T_{123\dots (N-1)|pN}(u-(N-2)i)
\eeq
where
\beq
	k=\{k_1,k_2\},\ \ l=\{l_1,l_2,l_3\},\ \dots, \ p=\{p_1,p_2,\dots,p_{N-2}\}
\eeq
and the sum runs over the values of indices $j,k_n,l_n,\dots,p_n$ from $1$ to $N-1$ with
\beq
	k_1<k_2,\ \ l_1<l_2<l_3,\ \dots,\ p_1<p_2<\dots <p_{N-2}\ .
\eeq
The main ingredient here is the pattern of indices in the expression under the sum, which for $SU(3)$,
$SU(4)$, $SU(5)$ reads
\beq
	jj, j\color{black}k\color{black}j\color{black}k\color{black},\
	jkj\color{black}l\color{black}k\color{black}l\color{black}
\eeq
 and so on -- we see that each time a new index
appears both before and after the last index of the previous expression. This pattern ensures in particular that all minors in \eq{BqN} have the appropriate number of indices. In appendix \ref{sec:appB} we present explicit expressions for $B$ corresponding to $N\leq 5$.

We also see that $B$ is a polynomial of degree $1+2+\dots+(N-1)=\frac{N(N-1)}{2}$ in the elements $T_{ij}$, as it should be according to the known classical result \eq{BclN}. For $N\leq 6$ we have also checked that in the classical limit our result precisely reproduces the classical expression \eq{BclN} (up to an overall sign which is irrelevant). This matching involves various cancellations which are certainly not obvious from the form of our result, and is already a highly nontrivial test of our conjecture.

Like in the lower rank cases, it is important to remove degeneracies by constructing the operator $B^{\rm good}$ from this operator $B$. We again use the similarity transformation \eq{Tg4} with a $N \times N$ matrix $K$ which we can take to be generic or specialize to an upper triangular $K$ with all elements equal to $0$ or $1$, like in \eq{Tg4}. Then we expect that the operator $B^{\rm good}(u)$ will provide a commutative family,
\beq
\label{BgcN}
	[B^{\rm good}(u),B^{\rm good}(v)]=0\ ,
\eeq
and will again generate the eigenstates of the transfer matrix by acting on the vacuum,
\beq
\label{B0N}
	\ket{\Psi}=B^{\rm good}(u_1)\dots B^{\rm good}(u_M)\ket{0}
\eeq
where $u_k$ are the momentum-carrying roots.

 Moreover, we expect that like in the $SU(3)$ and $SU(2)$ cases one can perform any duality transformation in the nested Bethe ansatz equations and then use the dual sets of momentum-carrying roots as arguments of $B^{\rm good}$ to build the states (and acting with $B^{\rm good}$ on the corresponding reference state instead of $\ket{0}$).

We have checked commutativity \eq{BgcN} and the construction of various states in the $SU(5)$ case for the first several values of $L$. Moreover for $L=2$ we explicitly verified with generic $K$ that our conjecture provides all the 25 eigenstates, going one by one through each solution of the rather involved Bethe equations which in particular contain four types of Bethe roots. In combination with the compact form of our operator $B$, these confirmations are hardly an accident and we believe they provide strong evidence for our proposal.

\subsection{Free parameters in the construction}
\label{sec:anyK}

It is important to reiterate that rather than just one operator $B^{\rm good}$, we can build a family of such operators by taking a generic $N\times N$ matrix $K$ in the expression \eq{Tg4} for the improved monodromy matrix $T^{\rm good}$, which reads
\beq
	T^{\rm good}=K^{-1}TK\ .
\eeq
Even with generic $K$, we observed in many examples that the operator $B^{\rm good}$ constructed from this $T^{\rm good}$ retains all the key properties such as commutativity and the ability to create the states. Without loss of generality we can impose $\det K=1$, as a scalar $K$ will not change $T$ at all. This leaves $N^2-1$ complex parameters in $K$. However, not all $K$'s lead to distinct operators $B^{\rm good}$. Based on explicit examples we considered, we expect that $B^{\rm good}$ will generically contain only $N$ free parameters (one of which corresponds to a trivial rescaling $B^{\rm good}\to {\rm const}\cdot B^{\rm good}$). We have checked this for the first several values of $L$ when $N\leq 4$ by considering the effect on $B^{\rm good}$ of small variation of the matrix $K$ around some generic matrix. This extra freedom may be useful in some applications and it would be interesting to better understand its role.

Let us mention that at the classical level one can easily introduce $N$ parameters into the expression for $B$ given in \eq{BclN}, namely can consider (see e.g. \cite{Diener:1994xt})\footnote{We are grateful to D. Medina and M. Heinze for related discussions.}
\beq
\label{BclNg}
	B=\sum_{i_1,\dots, i_{N}=1}^{N}
	\epsilon_{i_1\dots i_{N}} \alpha_{i_1} (T\alpha)_{i_2} (T^2 \alpha)_{i_3}\dots (T^{N-1}\alpha)_{i_{N}}\ ,
\eeq
where $\alpha=(\alpha_1,\dots,\alpha_N)$ is a constant vector. The original expression \eq{BclN} is recovered by setting $\alpha_i=\delta_{iN}$. Different choices of $\alpha$ correspond to different normalizations of the Baker-Akhiezer function which plays an important role in the SoV program for classical models (see \cite{Diener:1994xt,Sklyanin:1995bm} for details). Nicely, at the classical level the family of functions \eq{BclNg} parameterized by $\alpha_i$ is exactly the same as the family obtained in our approach which involves free parameters in the matrix $K$. Namely, by starting from the expression \eq{BclN} and replacing in it the matrix $T$ by $T^{\rm good}=K^{-1}TK$ one obtains precisely \eq{BclNg} in which the parameters $\alpha_i$ are given by\footnote{We have checked this for $N=2,3,4$.}
\beq
	\alpha_i=K_{iN},\ \ \ i=1,\dots,N\ 
\eeq
and we also assume that $\det K=1$. Thus at the classical level the $N$ parameters in our operator $B^{\rm good}$ correspond to the $N$-parametric freedom of choosing the normalization of the Baker-Akhiezer function. At the same time, this identification is only valid classically, and it would be important to clarify the algebraic meaning of the free parameters in our construction in the quantum case.

\subsection{Eigenvalues of separated variables}

It is natural to conjecture that the operator $B^{\rm good}$ we have constructed should also provide separated variables for the $SU(N)$ case. We expect all the main features to be in direct analogy with the $SU(2)$ and $SU(3)$ cases, and here we will briefly describe how they generalize to $SU(N)$.

In the $SU(N)$ case the operator $B^{\rm good}$ is a polynomial of degree $N(N-1)/2$ in $u$. We again define the operators $x_{n,a}$ as its zeros, with $n=1,\dots, L$ and $a=1,\dots,N(N-1)/2$. Then the construction of eigenstates \eq{B0N} implies that in the appropriately normalized left eigenbasis of $x_{n,a}$ (whose eigenvalues we denote as $\rx_{n,a}$) the wavefunction will factorize as
\beq
	\bra{\rx_{1,1},\dots,\rx_{L,N(N-1)/2}}\Psi\rangle=
	\prod_{n=1}^{L}\prod_{a=1}^{N(N-1)/2} (-1)^MQ_1(\rx_{n,a})
\eeq
where $Q_1(u)=\prod_{j=1}^M(u-u_j)$ is the Baxter function associated to momentum-carrying Bethe roots. Thus it is important to understand the spectrum of the $x_{n,a}$ operators, which can be found form the zeros of eigenvalues of $B^{\rm good}(u)$.

It is natural to associate $x_{n,a}$ with site $n$ of the chain. As before, we found that $B^{\rm good}$ contains a simple scalar factor multiplying a nontrivial matrix piece.
For $L=1$ the scalar factor is\footnote{There is also a $u$-independent prefactor which contains elements of the $K$ matrix, we do not write it here for brevity.}
\beqa
\label{exN}
	\[u-\(\theta_1+\frac{3i}{2}\)\]^{N-2}\[u-\(\theta_1+\frac{5i}{2}\)\]^{N-3}\dots
	\[u-\(\theta_1+\frac{(2N-3)i}{2}\)\]\
\eeqa
so we can choose all of the $x_{1,a}$ except $N-1$ of them to act as scalars with eigenvalues corresponding to zeros of \eq{exN}. The remaining $N-1$ of the $x_{1,a}$ have eigenvalues $\theta_1\pm \frac{i}{2}$, and the \textit{unordered} set of these eigenvalues identifies the eigenstate. We can therefore label the eigenstates by the pattern of $\pm$ signs in these eigenvalues, and all possible patterns are realized, giving $N$ states
\beq
	\ket{(+++\dots+)},\ \ket{(-++\dots+)},\ \ket{(--+\dots+)},\ \dots,\  \ket{(---\dots-)}
\eeq
which form a basis in the Hilbert space. For higher $L$ we expect that again eigenvalues are trivially combined between different sites, like in the $SU(3)$ case (see \eq{eig3}).


We have checked that the spectrum of $B$ has the form described above at least for the first several $L$'s in the $SU(4)$ and $SU(5)$ cases. This simple pattern of eigenvalues can also be viewed as yet another nontrivial test of our proposal \eq{BqN} for the $B$ operator in the $SU(N)$ case.

\section{Conclusions and future directions}

In this paper we put forward a new construction of eigenstates for $SU(N)$ integrable spin chains. It has a highly compact form which closely parallells the $SU(2)$ case, involving only a single operator $B(u)$ which is directly related to Sklyanin's SoV approach. We would like to emphasize that even for an arbitrary $SU(N)$ group the states are built using just one operator, rather than going through many levels of nesting in the usual algebraic Bethe ansatz.

Our proposal is supported by highly nontrivial analytic and numerical tests, leaving little doubt in its correctness. The simple pattern of eigenvalues of the separated variables that we observed also gives extra support to our conjecture. Our results also provide one of the very few concrete examples of the SoV program at work beyond the most-studied $SU(2)$ and $SU(3)$ cases.



Let us list several open problems and interesting directions for future work.

\begin{itemize}

\item One of the main motivations for us was the potential to apply the SoV program for computing 3-point correlators in planar $\cN=4$ supersymmetric Yang-Mills theory (SYM) and its dual string theory. The SoV has been already used successfully in this AdS/CFT context \cite{Kazama:2016cfl,Jiang:2015lda,Kazama:2015iua,Kazama:2014sxa,Kazama:2013qsa,Sobko:2013ema} (see also \cite{Belitsky:2014rba}), but its application has been restricted to essentially rank one sectors\footnote{See \cite{Foda:2013nua} for some direct calculations in the $SU(3)$ sector.}. We hope that our results will facilitate further progress, especially given that the $Q-$functions in $\cN=4$ SYM are available \textit{at any coupling} via the Quantum Spectral Curve proposed in \cite{Gromov:2013pga}. In view of the rather mysterious simplicity we observed for $SU(N)$ spin chains, one can hope for hidden simplifications in the $\cN=4$ theory and perhaps eventually obtain a framework allowing to access arbitrary correlators at finite coupling (despite impressive recent progress \cite{Basso:2015zoa}, this goal is far from having been accomplished). Let us also mention that the power of SoV in quantum field theory has already been demonstrated in various settings, see e.g. \cite{Smirnov:1998kv}.

\item It would be highly interesting to extend our construction to more general integrable models. This includes models based on trigonometric or elliptic R-matrices, spin chains with arbitrary representation of $SU(N)$ at each site and Gaudin-type models. We hope that it should also apply to various boundary problems, for example it would be interesting to study the interplay of our methods with the off-diagonal Bethe ansatz \cite{Cao:2013cja,Cao:2013nza}.

\item In many ways our construction is much simpler than the conventional nested Bethe ansatz, so it would be very interesting to prove it rigorously. One way to achieve this would be to derive concise commutation relations between the transfer matrix and the operator $B^{\rm good}$ which creates the states (even though it is a challenging task, similar calculations were done in e.g. \cite{Slavnov:2016nhu}). Knowing the commutation relations would also open the way to computing form factors for which only partial results are available in higher rank cases. It would also be interesting to try to obtain a proof using the variables canonically conjugated to the separated variables $x_n$, following an approach similar to the one we used to prove the construction for $SU(2)$ with a generic matrix $K$ (see Appendix \ref{app:proofsu2}). 

\item Deriving a compact expression for Sklyanin's measure in separated variables would further facilitate calculation of various observables for higher rank spin chains. This problem is particularly interesting for non-compact $sl(n)$ spin chains with infinite dimensional representations at each site, where partial results for the measure and expressions for Q-operators are available \cite{Derkachov:2003qb,Derkachov:2010qe}.

\item As our construction has a rather compact form compared to many of the other approaches, we hope it could be useful in attacking the challenging question of extending the celebrated Slavnov's determinant result for scalar products beyond the $SU(2)$ case.

\item It is curious that the \textit{same} operator $B^{\rm good}$ allows to build the states just as well using a \textit{dual} set of momentum-carrying Bethe roots, obtained via a duality transformation in the nested Bethe equations. Together with the presence of free parameters in $B^{\rm good}$ (as discussed in the end of section 5.2), this fact shows a surprising versatility of our construction whose implications remain to be understood.

\item An interesting question is to better understand the algebraic structure of our operator $B^{\rm good}$  for $SU(N)$, e.g. it might be possible to write it in terms of monodromy matrices in antisymmetric representations like in the $SU(3)$ case (see \eq{BU}). Moreover, one can consider the operator built like $B^{\rm good}$ but using the elements of such a monodromy matrix instead of those of the usual $T(u)$. We hope that the resulting operator may allow to generalize our construction to spin chains with other representations at each site.

\item While we presented an explicit result for the $B$ operator, it would be interesting to also build for $SU(N)$ the $A$ operator of Sklyanin \cite{Sklyanin:1992sm}, which provides variables canonically conjugated to the separated variables $x_n$.

\item In our construction it is important that all twists and inhomogeneities are switched on. Taking the limit corresponding to the homogenous periodic XXX chain is nontrivial, but should be possible to accomplish.

\item As the Q-functions feature prominently in our approach, it would be natural to look for links with the explicit construction of Q-operators from \cite{Frassek:2011aa}.

\item Surprisingly, the SoV approach is completely undeveloped even for the simplest \textit{supersymmetric} chains based on the $su(1|1)$ superalgebra. It would be of great interest to generalize our conjecture to super-spin chains, for which various ways of constructing the eigenstates have been explored recently in \cite{Belliard:2008di,Pakuliak:2016bhc} (see also \cite{Hutsalyuk:2016jwh,Hutsalyuk:2016yii,Slavnov:2016nhu}).

\item Finally, we hope that our results for the $SU(N)$ case may help to shed light on algebraic structures underlying the SoV approach, which has close links with deep subjects in mathematics such as the Langlands correspondence \cite{Frenkel:1995zp,Chervov:2006xk,Frenkel:2015rda}. It would be also interesting to explore possible relations with Talalaev's quantum spectral curve and the Manin matrices approach \cite{Talalaev:2004qi,Chervov:2007bb} as well as with classical/quantum and spectral dualities in higher rank integrable models (see e.g. \cite{Alexandrov:2011aa,Mironov:2012ba,Gorsky:2013xba,MTV1,Gorsky:1995zq}).

\end{itemize}

\section*{Acknowledgements}

We are grateful to J.~Caetano, M.~Heinze, Y.~Jiang, V.~Kazakov, S.~Komatsu, I.~Kostov,  A.~Liashyk, D.~Medina, D.~Serban, N.~Slavnov, F.~Smirnov, E.~Sobko, S.~Valatka, B.~Vicedo, J.-B.~Wu, K.~Zarembo and Y.~Zenkevich for discussions and helpful comments, and to S.~Valatka for collaboration at an early stage of this project. We are especially grateful to I.~Kostov and D.~Serban for many discussions. We also thank J.~Caetano and Y.~Jiang for sharing some of their Mathematica code. We would like to thank Ecole de Physique des Houches and Humboldt University Berlin, where a part of this work was done, for hospitality. The research leading to these results has received funding from the People Programme
(Marie Curie Actions) of the European Union's Seventh Framework Programme FP7/2007-
2013/ under REA Grant Agreement No 317089 (GATIS), from
European Research Council (Programme “Ideas” ERC-
2012-AdG 320769 AdS-CFT-solvable) and from the ANR
grant StrongInt (BLANC- SIMI- 4-2011).
This work was also supported by the grant ``Exact Results in Gauge and String Theories'' from the Knut and Alice Wallenberg foundation. We wish to thank
STFC for support from Consolidated
grant number ST/J002798/1.

\appendix

\section{Proving the construction of eigenstates for a generic matrix $K$ in the $SU(2)$ case}
\label{app:proofsu2}

In this section we present a proof of the fact (discussed in section \ref{sec:su2impl}) that for an $SU(2)$ spin chain the operator $B^{\rm good}$ generates the eigenstates as
\beq
	\ket{\Psi}=B^{\rm good}(u_1)\dots B^{\rm good}(u_M)\ket{0}
\eeq
even if it is constructed using a generic matrix $K$. Let us remind that this matrix is used to build an improved monodromy matrix $T^{\rm good}$ from the original $T$ (see \eq{Tgdef}),
\beq
\label{TKapp}
	T^{\rm good}=K^{-1}TK\ .
\eeq
We will focus on the spin chain with fundamental representation at each site. The only assumption we make is that the spectrum of eigenvalues of the separated variables $x_k$ is $\theta_k\pm i/2$ as discussed in the main text\footnote{This statement likely can also be proven rigorously along the lines of Appendix A in \cite{Kazama:2013rya}.}, and their eigenstates form a basis in the Hilbert space labelled as
$\ket{\rx_1,\dots,\rx_L}$. For other representations of $SU(2)$ the spectrum would be different but we expect a similar proof should work.

The main problem for generic $K$ is that the vacuum $\ket{0}$ is not an eigenstate of $A^{\rm good}(u)$ and $D^{\rm good}(u)$, so the usual proof breaks down. Our strategy is to follow instead the SOV approach and use the variables canonically conjugated to $x_k$. Eventually we will arrive at the Baxter equation which we know is satisfied by the Q-function
\beq
	Q_1(u)=\prod_{k=1}^M(u-u_k)\ ,
\eeq
which will complete the proof.

Roughly speaking the variables conjugated to $x_k$ are given by $A^{\rm good}(u)$ and $D^{\rm good}(u)$ evaluated at $u=x_k$, but one should be careful with operator ordering \cite{Sklyanin:1995bm}. Expanding these operators as
\beq
	A^{\rm good}(u)=\sum_{n=0}^LA_nu^n\ ,\ \ D^{\rm good}(u)=\sum_{n=0}^LD_nu^n,
\eeq
we define
\beq
\label{Aord}
	A^{\rm good}(x_k)=\sum_{n=0}^L(x_k)^nA_n\ ,\ \ 
	D^{\rm good}(x_k)=\sum_{n=0}^L(x_k)^nD_n\ .
\eeq
Then from the RTT commutation relations it follows that \cite{Sklyanin:1995bm} 
\beq
	A^{\rm good}(x_k)x_n=(x_n-i\delta_{kn}) A^{\rm good}(x_k),\ \ 
	D^{\rm good}(x_k)x_n=(x_n+i\delta_{kn}) D^{\rm good}(x_k)
\eeq
(note that $T^{\rm good}$ satisfies the same RTT algebra as the original $T$). This means that $A^{\rm good}(x_k)$ and $D^{\rm good}(x_k)$ act as raising and lowering operators\footnote{Therefore e.g. $\log \(A^{\rm good}(x_k)\)$ would have canonical commutation relations with the $x_n$ operators.}, so we have
\beq
\label{F1k1}
	\bra{\rx_1,\dots,\rx_k,\dots,\rx_L}A^{\rm good}(x_k)=
	F_{1k}(\rx)\bra{\rx_1,\dots,\rx_k-i,\dots,\rx_L}\ ,
\eeq
\beq
\label{F2k1}
	\bra{\rx_1,\dots,\rx_k,\dots,\rx_L}D^{\rm good}(x_k)=
	F_{2k}(\rx)\bra{\rx_1,\dots,\rx_k+i,\dots,\rx_L}\ ,
\eeq
where $F_{1k}(\rx),F_{2k}(\rx)$ are some scalar coefficients. Let us explain how to fix them.

First, since there is no state with eigenvalue of $x_k$ equal to $\theta_k-3i/2$, we must have\footnote{For spin chains with other representations of $SU(2)$ it may be possible to lower the eigenvalue to $\theta_k-3i/2$ and even beyond. However at some point this process should terminate as otherwise we would get an infinite-dimensional Hilbert space. We hope therefore that a similar proof should work for other $SU(2)$ representations as well.}
\beq
\label{F10}
	F_{1k}(\rx)=0\  \ \text{when}\ \  \rx_k=\theta_k-i/2\ .
\eeq
Similarly,
\beq
	F_{2k}(\rx)=0\  \ \text{when}\ \  \rx_k=\theta_k+i/2\ .
\eeq
To compute the remaining nonzero coefficients let us consider the matrix element
\beq
\label{mel}
	{\cal M}=\bra{\rx_1,\dots,\theta_k+i/2,\dots,\rx_L}\(A^{\rm good}(x_k)+D^{\rm good}(x_k)\)\ket{0}
\eeq
i.e. we chose $\rx_k=\theta_k+i/2$. On the one hand, $D^{\rm good}(x_k)$ annihilates this bra vector, so using \eq{F1k1} and our normalization
\beq
	\bra{\rx_1,\dots,\rx_L}0\rangle=1\ ,
\eeq
we see that the matrix element \eq{mel} is given by
\beq
\label{MF1}
	{\cal M}=F_{1k}(\rx_1,\dots,\theta_k+i/2,\dots,\rx_L)\ .
\eeq
On the other hand, due to the ordering in \eq{Aord} we can act with operators $x_k$ directly on the bra state, so the operator $x_k$ in the argument of $A^{\rm good}$ and $D^{\rm good}$ can be replaced by the constant $\theta_k+i/2$. While $\ket{0}$ is not an eigenvector of $A^{\rm good}$ or $D^{\rm good}$, it is still an eigenvector of $A^{\rm good}(\theta_k+i/2) + D^{\rm good}(\theta_k+i/2)$, because the transformation \eq{TKapp} does not change the trace of the monodromy matrix. The eigenvalue is given by
\beq
	\(A^{\rm good}(u) + D^{\rm good}(u)\)\ket{0}=
	\(\lambda_1Q_\theta^+(u)+\lambda_2Q_\theta^-(u)\)\ket{0}\ ,
\eeq
and substituting $u=\theta_k+i/2$ into this expression we see that the term with $\lambda_2$ drops out. This means that the matrix element \eq{mel} is equal to $\lambda_1Q_\theta^+(\theta_k+i/2)$. In combination with \eq{MF1} this finally gives
\beq
	F_{1k}(\rx_1,\dots,\theta_k+i/2,\dots,\rx_L)=\lambda_1Q_\theta^+(\theta_k+i/2)\ .
\eeq
We can write this result and \eq{F10} in a uniform way as
\beq
\label{F1kr}
	F_{1k}(\rx)=\lambda_1Q_\theta^+(\rx_k)\ .
\eeq
Similarly, one can show that
\beq
\label{F2kr}
	F_{2k}(\rx)=\lambda_2Q_\theta^-(\rx_k)\ .
\eeq

Having found the coefficients $F_{1k},F_{2k}$ it is now rather easy to complete the proof. As the transfer matrix is a polynomial in $u$, it is enough to show that $\ket{\Psi}$ is its eigenstate at $L$ distinct values of $u$. We will consider $u=\theta_k\pm i/2$ for all $k$ which gives more than enough points.

Namely, let us evaluate the scalar product
\beq
\label{Sdef}
	{\cal S}=\bra{\rx}\(A^{\rm good}(x_k) + D^{\rm good}(x_k)\)\ket{\Psi}\ 
\eeq
in two different ways like we just did for the case when $\ket{\Psi}$ is simply $\ket{0}$. First, due to the ordering in \eq{Aord} we can again replace the operator $x_k$ by its eigenvalue $\rx_k$, so that
\beq
\label{Sr}
	{\cal S}=\bra{\rx}\(A^{\rm good}(\rx_k) + D^{\rm good}(\rx_k)\)\ket{\Psi}\ .
\eeq
The operator appearing here is precisely the transfer matrix at $u=\theta_k\pm i/2$. Alternatively, we can use the fact that $A^{\rm good}(x_k)$ and $D^{\rm good}(x_k)$ act as shift operator on $\bra{\rx}$, so that
\beq
	{\cal S}=\lambda_1Q_\theta^+(\rx_k)\bra{\rx_1,\dots,\rx_k-i,\dots,\rx_L}\Psi\rangle
	+ \lambda_2Q_\theta^-(\rx_k)\bra{\rx_1,\dots,\rx_k+i,\dots,\rx_L}\Psi\rangle
\eeq
where we used \eq{F1k1}, \eq{F2k1}, \eq{F1kr}, \eq{F2kr}. The scalar products appearing here are just products of Q-functions, so using \eq{facsu2} we get
\beq
	{\cal S}=\[\lambda_1Q_\theta^+(\rx_k)Q_1^{--}(\rx_k)
	+ \lambda_2Q_\theta^-(\rx_k)Q_1^{++}(\rx_k)\](-1)^{ML}\prod_{j\neq k}^LQ_j(\rx_j)\ .
\eeq
Inside the square brackets we recognize part of the Baxter equation! The equation itself reads
\beq
	\tau(u)Q_1(u)=\lambda_1Q_\theta^+(u)Q_1^{--}(u)
	+ \lambda_2Q_\theta^-(u)Q_1^{++}(u)\ ,
\eeq
where $\tau(u)$ is the transfer matrix eigenvalue. This means that
\beq
	{\cal S}=\tau(\rx_k)Q_1(\rx_k)(-1)^{ML}\prod_{j\neq k}^LQ_j(\rx_j)\ ,
\eeq
which can also be written as
\beq
	{\cal S}=\bra{\rx}\tau(\rx_k)\ket{\Psi}\ .
\eeq
Comparing the last expression with \eq{Sr} we see that for all ${\bra{\rx}}$ we have
\beq
	\bra{\rx}\(A^{\rm good}(\rx_k) + D^{\rm good}(\rx_k)\)\ket{\Psi}=\bra{\rx}\tau(\rx_k)\ket{\Psi}\ .
\eeq
Since the states $\bra{\rx}$ form a complete basis, we get
\beq
	\(A^{\rm good}(\rx_k) + D^{\rm good}(\rx_k)\)\ket{\Psi}=\tau(\rx_k)\ket{\Psi}\ .
\eeq
In other words, $\ket{\Psi}$ is an eigenstate of the transfer matrix when $u=\theta_k\pm i/2$. The polynomiality of the transfer matrix now guarantees that $\ket{\Psi}$ is its eigenstate for any $u$, thus completing the proof.

\section{Derivation of the norm for $SU(2)$ spin chains}

\label{app:norm}

Here we give technical details on the derivation of the norm of a spin chain eigenstate in the $SU(2)$ case, mentioned in the main text in section \ref{norm1}.

In order to find the norm of a state it is convenient to consider it together with another one, corresponding to the dual set of Bethe roots.
Below we use the following notation:
\beq
\{u_{k=1\dots M}\}\;\;\;\;  \textrm{and}\;\;\;\; \{v_{k=1\dots L-M}\},
\eeq
are the original set of Bethe roots and the dual roots, satisfying the Bethe equations, while
\beq
\label{psi12std}
	\ket{\Psi_1^{std}}=B(u_1)\dots B(u_M)\ket{0} \;\;\;\; \textrm{and} \;\;\;\;
	\ket{\Psi_2^{std}}=C(v_1)\dots C(v_{L-M})\ket{0'}
\eeq
are the two states we consider, with $\ket{0}$ and $\ket{0'}$ being the usual vacuum and the dual vacuum (defined in \eq{def0}, \eq{su20p}). By $N_1,N_2$ we denote the norms of the states \eq{psi12std}. We will study the case when the twists satisfy
\beq
	\lambda_1=1/\lambda_2=1/\lambda_1^*\ ,
\eeq
and accordingly the set of Bethe roots $\{u_k\}$ is invariant under complex conjugation (the same is true for dual roots $\{v_k\}$).

In the first subsection below we derive an expression for the ratio of norms of these two states via SoV, and in the second one we obtain an expression for their product via the usual Gaudin's formula. In the second part we compute as an intermediate step the variation of Q-functions with respect to the inhomogeneities. Combining the ratio and the product of the norms, we obtain the final result for the individual norm $N_1$, given in \eq{N1app}.

\subsection{Ratio of the norms}
\label{subsec:Ratio}

To derive the ratio of the norms we will make use of construction of the states via the operator $B^{\rm good}$ which provides separated variables for our spin chain. When building this operator it is convenient to choose the matrix $K$ in \eq{Tgdef} to be of the form
\beq
	K=
	\begin{pmatrix}
	\frac{1}{\sqrt{\alpha}}\ \ &-\frac{1}{\sqrt{\alpha}}
	\\
	0&\sqrt{\alpha}
	\end{pmatrix}
\eeq
and to multiply $B^{\rm good}$ by an extra overall factor of $1/(\lambda_2-\lambda_1)$ as in \eq{Balp}, so that we get
\beq
	B^{\rm good}=\frac{\alpha B-1/\alpha\; C-A+D}{\lambda_2-\lambda_1} \ .
\eeq
Here $\alpha$ is an unfixed parameter which we retain for convenience.
Consider the left eigenvector of $B^{\rm good}$ with all eigenvalues of the form $\theta_k+i/2$, denoted as $\bra{+\dots++}$. We have observed that
\beq\label{NullNull}
	\bra{+\dots++} 0'\rangle=\(\frac{\lambda_1}{\lambda_2\alpha}\)^L
	\bra{+\dots++} 0 \rangle \ .
\eeq
We have checked this curious fact for the first several $L$'s, and leave a general proof for the future.

Let us introduce two states analogous to $\ket{\Psi_1^{std}}$ and  $\ket{\Psi_2^{std}}$ but created with the operator $B^{\rm good}$ instead of $B$ and $C$:
\beq
\label{psi12}
	\ket{\Psi_1}=B^{\rm good}(u_1)\dots B^{\rm good}(u_L)\ket{0},\ \
	\ket{\Psi_2}=B^{\rm good}(v_1)\dots B^{\rm good}(v_{L-M})\ket{0'}\ .
\eeq
As we explained in section \ref{sec:su2impl}, the same operator $B^{\rm good}(u)$ allows to construct the states using either usual Bethe roots or dual Bethe roots. This means that the two states \eq{psi12} are both eigenvectors of $T(u)$ with the same eigenvalue, so they should be collinear\footnote{We recall that with for nontrivial twists $\lambda_1,\lambda_2$ all eigenvalues of the transfer matrix have multiplicity one.}, i.e. for some $c$
\beq
	\ket{\Psi_1}=c \ket{\Psi_2}\ .
\eeq
Let us find the constant $c$. To this end, consider the scalar product $\bra{+\dots++}\Psi_1\rangle$. Since $\ket{\Psi}$ was created with the operator $B^{\rm good}(u)$ and $\bra{+\dots++}$ is an eigenstate of $B^{\rm good}(u)$, we get
\beq
	\bra{+\dots++}\Psi_1\rangle=\bra{+\dots++}0\rangle\prod\limits_{k=1}^L(-1)^{M}Q_1^+(\theta_k)
	\ .
\eeq
Similarly,
\beq
	\bra{+\dots++}\Psi_2\rangle=\bra{+\dots++}0'\rangle\prod\limits_{k=1}^L(-1)^{L-M}Q_2^+(\theta_k)\ .
\eeq
Using the relation \eqref{NullNull} we get
\beq
	\bra{+\dots++}\Psi_2\rangle=\bra{+\dots++}0\rangle\(\frac{\lambda_1}{\lambda_2\alpha}\)^L\prod\limits_{k=1}^L(-1)^{L-M}Q_2^+(\theta_k)\ ,
\eeq
and thus
\beq
c=\(\frac{\alpha\lambda_2}{\lambda_1}\)^L
\prod\limits_{k=1}^L (-1)^{L}\frac{Q_1^+(\theta_k)}{Q_2^+(\theta_k)}
\ .
\eeq
We also observed that
\beq
	\ket{\Psi_1}=\(\frac{\alpha}{\lambda_2-\lambda_1}\)^{M} \ket{\Psi_1^{std}},
	\ \
	\ket{\Psi_2}=\(-\frac{1/\alpha}{\lambda_2-\lambda_1}\)^{L-M} \ket{\Psi_2^{std}} \ .
\eeq
We have checked this identity for the first several values of $L$ and postpone a rigorous proof to later work.
 These relations allow us to express the ratio of norms of $\ket{\Psi_1^{std}}$ and $\ket{\Psi_2^{std}}$ through the ratio of norms of $\ket{\Psi_1}$ and $\ket{\Psi_2}$, which is equal to $|c|$, so finally we get the simple result
\beq\label{ratio}
	N_1/N_2=\left|
	(\lambda_2-\lambda_1)^{2M-L}\prod\limits_{k=1}^L\frac{ Q_1^+(\theta_k)}{ Q_2^+(\theta_k)}
	\right|
	\ .
\eeq

\subsection{Product of the norms}
\label{subsec:Product}

Let us now calculate the product of norms of the states $\ket{\Psi_1^{std}}$ and $\ket{\Psi_2^{std}}$ in terms of the Q-functions. Each of the norms is given by the Gaudin formula \cite{Korepin:1982gg},
\beq\label{defN1}
N_1^2=(i\lambda_1\lambda_2)^M\prod_{\substack{k=1\dots M \\ n=1\dots L}}\((u_k-\theta_n)^2+\frac{1}{4}\)\prod_{j< k}\(1+\frac{1}{(u_j-u_k)^2}\)\det\frac{\d f_k^{(1)}}{\d u_l}\ ,
\eeq
\beq\label{defN2}
N_2^2=(i\lambda_1\lambda_2)^{L-M}\prod_{\substack{k=1\dots L-M \\ n=1\dots L}}\((v_k-\theta_n)^2+\frac{1}{4}\)\prod_{j< k}\(1+\frac{1}{(v_j-v_k)^2}\)\det\frac{\d f_k^{(2)}}{\d v_l}\ ,
\eeq
where
\beq
f_k^{(1)}=\sum\limits_{i=1}^L \ln\(\frac{u_k-\theta_i+i/2}{u_k-\theta_i-i/2}\)+\sum\limits_{i=1}^M
\ln\(\frac{u_k-u_i-i}{u_k-u_i+i}\)\ ,
\eeq
\beq
f_k^{(2)}=\sum\limits_{i=1}^L \ln\(\frac{v_k-\theta_i+i/2}{v_k-\theta_i-i/2}\)+\sum\limits_{i=1}^{L-M}\ln\(\frac{v_k-v_i-i}{v_k-v_i+i}\)\ .
\eeq
We see that in the product of these two norms, a product of the Jacobians will appear. The trick we will use is rewriting this product in a simpler form by switching between variables $\{u_j,v_j\}$ and $\{\theta_n\}$.


Namely, let us define a function
\beq
	F=\ln \frac{z^+}{z^-}\ ,\ 
\eeq
with
\beq
z=\frac{\lambda_2Q_1^+ Q_2^--\lambda_1Q_1^- Q_2^+}{(\lambda_2-\lambda_1)Q_\theta}\ .
\eeq
We will pack $\{u_i\}$ and $\{v_j\}$ into one vector $\{U_k\}=\{u_1,\dots,u_{M},v_1,\dots,v_{L-M}\}$ and denote $F_k\equiv F(U_k)$. Then
\beq
	F_k=-f^{(1)}_k+\ln \(-\frac{\lambda_2}{\lambda_1}\),\ \ k=1,\dots, M\ \ \ ,
\eeq
\beq
	F_k=-f^{(2)}_k+\ln \(-\frac{\lambda_1}{\lambda_2}\),\ \ k=M+1,\dots, L\ \ \ .
\eeq
Consider the Jacobian $\frac{\d F_k}{\d U_n}$. On the one hand, this matrix is block-diagonal, so
\beq\label{onehand}
\det \frac{\d F_k}{\d U_n}=(-1)^L\det \frac{\d f^{(1)}(u_k)}{\d u_n}\det \frac{\d f^{(2)}(v_k)}{\d v_n}\ .
\eeq
On the other hand, using the chain rule we have\footnote{Notice that if Bethe equations are satisfied, $F_k=0$.}
\beq\label{otherhand}
\det\frac{\d F_k}{\d U_n}=-\det\frac{\d F_k}{\d\theta_p}\det\frac{\d \theta_p}{\d U_n}\ .
\eeq
The first Jacobian in this product is easy to compute, and we get
\beq
\frac{\d F_k}{\d\theta_p}=R_k\frac{i}{(U_k-\theta_p)^2+\frac{1}{4}}\ ,
\eeq
where
\beq
R_k=
\begin{cases}
1,\;\;\;\;k=1\dots M\\
-1,\;\;\;\; k=M+1\dots L
\end{cases}\ .
\eeq
The second Jacobian in \eqref{otherhand} can be calculated using the QQ-relation
\beq\label{rel1}
\lambda_2 Q_1^+ Q_2^- - \lambda_1 Q_1^- Q_2^+=(\lambda_2-\lambda_1) Q_\theta \ .
\eeq
Indeed, let us vary one $\theta_l$ in \eqref{rel1} infinitesimally, then the solutions of Bethe equations $u_i$ and $v_i$ will vary as well, but the equation should still hold, so we get
\beqa
-\sum\limits_{k=1}^M \(\lambda_2\frac{Q_1^+Q_2^-}{u-u_k+i/2}-\lambda_1\frac{Q_1^-Q_2^+}{u-u_k-i/2}\)\delta u_k-
\\
\sum\limits_{k=1}^{L-M} \(\lambda_2\frac{Q_1^+Q_2^-}{u-v_k-i/2}-\lambda_1\frac{Q_1^-Q_2^+}{u-v_k+i/2}\)\delta v_k
=(\lambda_2-\lambda_1)\frac{\d Q_\theta}{\d \theta_l}\delta \theta_l \ .
\eeqa
Evaluating this equation on $u=\theta_n$ and noticing that \eq{rel1} implies that
\beq
\label{Q12pm}
	\lambda_2 Q_1^+(\theta_n)Q_2^-(\theta_n)=\lambda_1 Q_1^-(\theta_n)Q_2^+(\theta_n)
	\ ,
\eeq	
we get
\beq
	\sum_{k=1}^L R_k\frac{\delta U_k}{\delta \theta_l}
	i\frac{\lambda_2}{\lambda_2-\lambda_1}\frac{\d Q_{\theta}(\theta_n)}{\d \theta_n}
	\frac{Q_1^+(\theta_n)Q_2^-(\theta_n)}{(\theta_n-U_k)^2+1/4}
	=\delta_{ln}\ .
\eeq
Therefore,
\beq
\frac{\d \theta_m}{\d U_k}=R_k\frac{i \lambda_2}{\lambda_2-\lambda_1} \frac{Q^+_1(\theta_m)Q^-_2(\theta_m)}{\prod\limits_{n\ne m}(\theta_n-\theta_m)}\frac{i}{(U_k-\theta_m)^2+\frac{1}{4}} \ .
\eeq
Plugging this back into \eqref{otherhand} gives
\beq\label{FU}
\det \frac{\d F_k}{\d U_n}=\(\frac{i \lambda_2}{\lambda_2-\lambda_1}\)^L \frac{\prod\limits_m Q^+_1(\theta_m)Q^-_2(\theta_m)}{\prod\limits_{n< m}(\theta_n-\theta_m)^2}\(\det\frac{i}{(U_k-\theta_m)^2+\frac{1}{4}}\)^2.
\eeq
We can now multiply the starting expressions \eqref{defN1}, \eqref{defN2} for the norms and express the product $\det \frac{\d f^{(1)}(u_k)}{\d u_n}\det \frac{\d f^{(2)}(v_k)}{\d v_n}$ through \eqref{onehand} and \eqref{otherhand}. Finally we use \eqref{FU} to get an important intermediate result,
\beqa
(N_1 N_2)^2&=&(-1)^L\(\frac{i\lambda_2}{\lambda_2-\lambda_1}\)^{L} \prod_n Q_1^+(\theta_n)Q_1^-(\theta_n)Q_2^+(\theta_n)Q_2^-(\theta_n)\\
&\times& \prod\limits_{n<m}\(1+\frac{1}{(u_n-u_m)^2}\)\prod\limits_{i<j}\(1+\frac{1}{(v_i-v_j)^2}\)\\
&\times& \frac{\prod\limits_m Q^+_1(\theta_m)Q^-_2(\theta_m)}{\prod\limits_{n< m}(\theta_n-\theta_m)^2}\(\det\frac{i}{(U_k-\theta_m)^2+\frac{1}{4}}\)^2\ .
\eeqa

Now let us simplify this expression. First we can expand the determinant using the identity\footnote{This identity can be easily proven by writing the determinant as a sum over permutations and representing each matrix element as a sum of two simple fractions. Then after expansion of each product we get a sum over $2^n$ combination of signs. The order of the two sums (over permutations and over signs) can be switched, so the sums over permutations are reassembeld in $2^n$ determinants.}
\beq
\det\frac{i}{(U_k-\theta_m)^2+\frac{1}{4}}=\sum\limits_{\{s_i=\pm1\}}\prod\limits_{i=1}^Ls_i\det \frac{i}{U_k-\theta_m+\frac{i s_m}{2}}\ .
\eeq
Applying  the Cauchy determinant formula to each of the individual determinants, we obtain
\beqa
\nonumber
&&|N_1 N_2|^2=\left|\(\frac{1}{\lambda_2-\lambda_1}\)^{L}\prod\limits_{n<m}\(1+(u_n-u_m)^2\)\prod\limits_{i<j}\(1+(v_i-v_j)^2\)\prod_{m,n}(u_m-v_n)^2 \right|\times\\
&&\times\left| \frac{\(\prod\limits_m Q^+_1(\theta_m)Q^-_2(\theta_m)\)^{3}}{\prod\limits_{n< m}(\theta_n-\theta_m)^2}
\(\sum\limits_{\{s_i=\pm1\}}\prod\limits_{i=1}^Ls_i\frac{\prod\limits_{m<n}
\[\theta_n-\theta_m+i\frac{s_n}{2}-i\frac{s_m}{2}\]}{\prod\limits_m Q^{[s_m]}_1(\theta_m)Q^{[s_m]}_2(\theta_m)}\)^2\right|\nonumber\ .
\eeqa
Next, we notice that for any sets $\{u_k\}, \{v_k\}$ (regardless of Bethe equations)
\beqa\nonumber
&&\prod\limits_{n<m}\(1+(u_n-u_m)^2\)\prod\limits_{i<j}\(1+(v_i-v_j)^2\)
\prod_{m,n}(u_m-v_n)^2=
\\
&&=C\prod\limits_n Q_1^{++}(u_n) Q_2(u_n)\prod_m Q_1(v_m) Q_2^{--}(v_m)\nonumber\ ,
\eeqa
where the irrelevant constant $C$ is $i$ to some integer power.
Using that as a consequence of \eqref{rel1} we have
\beqa
\lambda_2 Q_1^{++}(u_k)Q_2(u_k)=(\lambda_2-\lambda_1)Q^+_\theta(u_k)\ , \\
\lambda_2 Q_1(v_k)Q_2^{--}(v_k)=(\lambda_2-\lambda_1)Q^-_\theta(v_k)\ ,
\eeqa
we get
\beqa\nonumber
&&\left|\prod\limits_{n<m}\(1+(u_n-u_m)^2\)\prod\limits_{i<j}\(1+(v_i-v_j)^2\)\prod_{m,n}(u_m-v_n)^2\right|
=
\\
&&=\left|\(\lambda_1-\lambda_2\)^{L}\prod\limits_{k=1}^M
Q_\theta^+(u_k)\prod\limits_{k=1}^{L-M}Q_\theta^-(v_k)\right|\nonumber\ ,
\eeqa
which after some cancellations and application of \eq{Q12pm} results in
\beq
	\left|N_1N_2\right|=
\left|\sum\limits_{\{s_i=\pm1\}}\prod\limits_{i=1}^Ls_i
{\prod\limits_{m<n}
\frac{\[\theta_n-\theta_m+i\frac{s_n}{2}-i\frac{s_m}{2}\]}{
{\theta_n-\theta_m}
}}{\prod\limits_m Q^{[-s_m]}_1(\theta_m)Q^{[-s_m]}_2(\theta_m)}\right|\ ,
\eeq
where we have taken into account that $\lambda_1=1/\lambda_2$. This is the final result for the product of the norms.

As a last step we combine this result with the formula \eqref{ratio} for the ratio of the norms, which gives the result for the norm of the usual transfer matrix eigenstate $\ket{\Psi_1^{std}}$ from \eq{psi12std} in terms of Q-functions only,
\beqa
\nn
	N_1&=&\left|
\sum\limits_{\{s_i=\pm1\}}\prod\limits_{i=1}^Ls_i{\prod\limits_{m<n}^L
\frac{\[\theta_n-\theta_m+i\frac{s_n}{2}-i\frac{s_m}{2}\]}{
{\theta_n-\theta_m}
}}{\prod\limits_{m=1}^L Q^{[-s_m]}_1(\theta_m)Q^{[-s_m]}_2(\theta_m)}
\right|^{1/2}\\ \label{N1app}
	&\times&
	\left|
	(\lambda_1-1/\lambda_1)^{2M-L}\;\prod\limits_{m=1}^L \frac{Q_1^+(\theta_m)}
	{Q_2^+(\theta_m)}
	\right|^{1/2}\ .
\eeqa

\section{The $B$ operator for $SU(N)$ with $N=2,3,4,5$}
\label{sec:appB}

Here we explicitly write the $B$ operator for spin chains with $SU(N)$ symmetry with $2\leq N\leq 5$. The general result for any $SU(N)$ is given in \eq{BqN} and explicit expressions are:
\beq
\begin{array}{crcl}\nn
	SU(2): & \ B(u)&=&T_{1|2}(u)\\ \nn \\ \nn
	SU(3): & \ B(u)&=&\sum\limits_{j}T_{j|3}(u)T_{12|j3}(u-i)\\ \nn \\ \nn
	SU(4): & \ B(u)&=&\sum\limits_{j,k}T_{j|4}(u)T_{k|j4}(u-i)T_{123|k4}(u-2i)\\ \nn \\ \nn
	SU(5): & \ B(u)&=&\sum\limits_{j,k,l}T_{j|5}(u)T_{k|j5}(u-i)T_{l|k5}(u-2i)T_{1234|l5}(u-3i)
\end{array}
\eeq
These formulas are written in terms of quantum minors defined in \eq{qmindef}. We used a shorthand notation
\beq
	k=\{k_1,k_2\},\ \ l=\{l_1,l_2,l_3\}
\eeq
and for $SU(N)$ the sum runs over the values of indices $j,k_n,l_n$ from $1$ to $N-1$ with
\beq
	k_1<k_2,\ \ l_1<l_2<l_3\ .
\eeq

\end{document}